\documentclass[letterpaper, twocolumn, 10pt]{article}

\usepackage{verbatim}
\usepackage[active,
  generate=synopsis,
  extract-env={},
  extract-cmd={section,subsection},
  extract-cmdline={synopsis}]{extract}
\begin{extract}
\usepackage{usenix-2020-09}

\usepackage{tikz}
\usepackage{amsmath}
\usepackage{pifont}
\usepackage{cleveref}

\usepackage{microtype}

\usepackage{todonotes} %
\usepackage{soul}

\usepackage{subfig}

\usepackage[sort&compress, numbers]{natbib}
\usepackage{xspace}
\usepackage{enumitem} %

\ifdefined\SHOWDELTA%

\newcommand{\revisionnumber}[1]{}
\else%

\newcommand{\revisionnumber}[1]{}
\fi%

\newcommand{\DONE}{}
\ifdefined\DONE%

\newcommand{\rk}[1]{}
\newcommand{\jm}[1]{}
\newcommand{\mj}[1]{}
\newcommand{\kr}[1]{}
\newcommand{\mm}[1]{}
\newcommand{\dk}[1]{}
\newcommand{\red}[1]{}
\newcommand{\commentbox}[2][noshadow]{}
\newcommand{\synopsis}[2][noshadow]{}
\else%

\newcommand{\rk}[1]{}
\newcommand{\jm}[1]{}
\newcommand{\mj}[1]{}
\newcommand{\kr}[1]{}
\newcommand{\mm}[1]{}
\newcommand{\dk}[1]{}
\newcommand{\red}[1]{{\color{red}\bf#1}}
\newcommand{\commentbox}[2][noshadow]{}

\newcommand{\synopsis}[2][noshadow]{}
\fi%
\newcommand{\hide}[1]{}

\usepackage{makecell}

\usepackage[plainruled, linesnumbered]{algorithm2e}
\usepackage{listings}
\usepackage{algpseudocode}
\SetAlCapSkip{1em}
\SetAlgoCaptionSeparator{:}

\usepackage{tabularx}
\usepackage{tabularray}
    \UseTblrLibrary{booktabs}
\usepackage{booktabs}
\usepackage[para,online,flushleft]{threeparttable}

\newcommand{\mcts}{MCTS}
\newcommand{\montecarlo}{Monte Carlo tree search}
\newcommand{\xss}{XSS}
\newcommand{\blindxss}{blind XSS}
\newcommand{\Ourscriptname}{Notification Script} %
\newcommand{\ourscriptname}{\MakeLowercase\Ourscriptname} %

\newcommand{\hookleft}{\raisebox{0ex}[0ex][0ex]{\textcolor{black}{\tiny\ensuremath{\hookleftarrow}}}}

\newcommand{\dcircle}[1]{\ding{\numexpr181 + #1}} %

\makeatletter
\newcommand{\currentfontsize}{\fontsize{\f@size}{\f@baselineskip}\selectfont}
\makeatother

\newcommand{\tablesize}{\footnotesize}
\newcommand{\codesize}{\scriptsize}
\newcommand{\inlinecodesize}{\currentfontsize}
\newcommand{\linksize}{\currentfontsize}

\usepackage{siunitx}
\usepackage{MnSymbol} %
\usepackage[frozencache]{minted}

\setminted{
	breaklines,
	escapeinside=||,
	style=xcode,
	numberblanklines=true,
	tabsize=2,
	linenos=true,
	frame=lines,
	fontsize=\codesize,
	framesep=1mm,
	xleftmargin=1em,
}

\usepackage[available,functional]{usenixbadges}

\definecolor{stringred}{HTML}{CC0000}

\setmintedinline{fontsize=\currentfontsize}

\newcommand{\htmlsquote}[1]{\PYG{l+s}{\PYGZsq{}}\PYG{}{\emph{#1}}\PYG{l+s}{\PYGZsq{}}} %

\newcommand{\htmldquote}[1]{\PYG{l+s}{\PYGZdq{}}\PYG{}{\emph{#1}}\PYG{l+s}{\PYGZdq{}}} %
\newcommand{\htmldquotesingle}[1]{\PYG{l+s}{\PYGZdq{}}\PYG{}} %

\newcommand{\inlinecode}[1]{{\inlinecodesize\texttt{#1}}} %
\newcommand{\code}[1]{\inlinecode{#1}} %

\newlist{questions}{enumerate}{2}
\setlist[questions,1]{label=\emph{RQ\arabic*:},ref=RQ\arabic*,leftmargin=1cm}

\makeatletter
\newcommand\etc{etc\@ifnextchar.{}{.\@}}
\makeatother

\makeatletter
\DeclareRobustCommand\onedot{\futurelet\@let@token\@onedot}
\def\@onedot{\ifx\@let@token.\else.\null\fi\xspace}

\def\eg{e.g\onedot}
\def\ie{i.e\onedot}

\def\etc{etc\onedot}

\makeatother

\newcommand{\linkstyle}[1]{{\linksize\nolinkurl{#1}}} %

\end{extract}
\usepackage{pdfpages}

\begin{document}
\pagenumbering{gobble}
\date{}

\title{\Large \bf Dancer in the Dark:\\ 
  Synthesizing and Evaluating Polyglots for Blind Cross-Site Scripting}

\author{
{\rm Robin Kirchner\thanks{Corresponding author}~\hspace{2px}\textsuperscript{\textdagger}, Jonas Möller\hspace{1px}\textsuperscript{\textdaggerdbl}, Marius Musch\textsuperscript{\textdagger}, David Klein\textsuperscript{\textdagger}, Konrad Rieck\hspace{1px}\textsuperscript{\textdaggerdbl}, Martin Johns\textsuperscript{\textdagger}}\\[0.5em]
\textsuperscript{\rm \textdagger} Technische Universität Braunschweig, Germany\\
\textsuperscript{\rm \textdaggerdbl} Technische Universität Berlin, Germany\\[0.5em]
\{robin.kirchner, m.musch, david.klein, m.johns\}@tu-braunschweig.de\\
\{jonas.moeller.1, rieck\}@tu-berlin.de
} %

\maketitle

\begin{abstract}
Cross-Site Scripting (XSS) is a prevalent and well known security problem in web applications. 
Numerous methods to automatically analyze and detect these vulnerabilities exist. 
However, all of these methods require that either code or feedback from the application is available to guide the detection process. 
In larger web applications, inputs can propagate from a frontend to an internal backend that provides no feedback to the outside. 
None of the previous approaches are applicable in this scenario, known as \emph{\blindxss{}} (BXSS).
In this paper, we address this problem and present the first comprehensive study on BXSS\@. 
As no feedback channel exists, we verify the presence of vulnerabilities through blind code execution. 
For this purpose, we develop a method for synthesizing \emph{polyglots}, small XSS payloads that execute in all common injection contexts. 
Seven of these polyglots are already sufficient to cover a state-of-the-art XSS testbed.
In a validation on real-world client-side vulnerabilities, we show that their XSS detection rate is on par with existing taint tracking approaches.
Based on these polyglots, we conduct a study of BXSS vulnerabilities on the Tranco Top \num{100000} websites. 
We discover \num{20} vulnerabilities in \num{18} web-based backend systems.
These findings demonstrate the efficacy of our detection approach and point at a largely unexplored attack surface in web security.
\end{abstract}

\section{Introduction}

Web applications are ubiquitous---so are their security problems.
In particular, the vulnerability class of Cross-Site Scripting (XSS) remains one of the major security issues on the Web.
Over the past decade, XSS has been studied extensively, with a focus on approaches to detect these flaws in existing applications.
This detection relies on the identification of insecure data flows from attacker-controlled inputs into security-sensitive sinks. 
The actual techniques to pursue this goal vary, but they all rely on the availability of observable evidence of such data flows.
This can be as simple as the re-identification of an input string in the application's HTML, in the case of manual bug hunting, and as sophisticated as fully automatic taint analysis.

This poses a problem for an often overlooked component of large web applications: \emph{the backend}. Similar to the use of web technologies in its public frontend, the same application often employs web interfaces for its internal backends.
These interfaces, inaccessible from the public internet behind a firewall, serve various internal application functions, including administration of the application, usage analysis, and back-office tasks such as supply chain management and accounting. 

Not different from public web applications, internal web tools are also susceptible to security problems in general and XSS in particular. 
As these backend systems process data generated by the application's public counterpart, there is a high likelihood of data from potentially malicious, attacker-supplied inputs flowing into the backend's web code. 
This situation opens the door for XSS attacks on internal application components.
However, current XSS detection techniques fall short in finding such flaws, as no evidence of problematic data flows can be obtained. All effects of the public-to-internal flows remain within the constraints of the internal systems.
Testing tools operating from the outside are literally ``blind'' to any side effects caused by injection attempts. 

In this paper, we present the first comprehensive study on \textit{Blind Cross-Site Scripting (BXSS)} vulnerabilities. As by definition \textit{no} direct channels from the affected system-under-test and the testing approach exist, the only option to verify the presence of a BXSS vulnerability is the injection and execution of JavaScript code. 
To this end, we present a method for synthesizing \emph{polyglots}: XSS payloads specifically designed for a multitude of injection contexts. More specifically, we use Monte Carlo tree search (MCTS) to synthesize a small set of seven polyglots, capable of executing code in all test cases of a state-of-the-art XSS testbed (see \Cref{sec:generation}). 

However, good performance on an artificial testbed does not necessarily result in similar properties in real-world settings. For this reason, we validate the generated polyglots on a set of recently uncovered client-side XSS vulnerabilities found in the Tranco Top \num{10000} websites. To do so, we compare the polyglots' capability to uncover vulnerabilities with the currently established state-of-the-art method of precise payload generation~\cite{StoPfiKaiLek+15,LekStoJoh13,MelDasShaBau+18,BenKleBarJoh21,KleBarBen+22,stock2017web}. In this experiment our blind polyglots prove to be highly competitive with precise exploit generation, through triggering 147 vulnerabilities compared to 145 found by the precise payload generation. Furthermore, the generated polyglots vastly outperform the well-known, manually crafted ``Ultimate Polyglot''~\cite{ultimate-polyglot} (see \Cref{sec:taint-validation} for details).

Finally, we present the---to the best of our knowledge---first study on BXSS in the wild. We conduct a large-scale analysis of the Tranco Top \num{100000} websites using our generated polyglots as testing payloads, uncovering~\num{20} instances of BXSS problems in internal systems. This study demonstrates that our testing approach is well suited to detect BXSS vulnerabilities in real-world websites, as well as, the existence of this vulnerability class (see \Cref{sec:bxss-study}) on popular websites.

\paragraph{Contribution.} In summary, we make the following contributions in this paper:
\begin{itemize}
\setlength\itemsep{-0.2em} %

\item We propose a method for automatically synthesizing a small set of XSS polyglots using Monte Carlo tree search, covering all currently known injection contexts. %

\item We show that the polyglots synthesized by our method achieve comparable performance to state-of-the-art vulnerability detection for client-side \xss{}. %

\item We design a methodology to detect backend XSS automatically and use it to conduct a large-scale evaluation on the \num{100000} most popular websites, finding \num{20}~BXSS vulnerabilities in \num{18} backend systems. %
\end{itemize}

\section{Blind Cross-Site Scripting}\label{sec:setting}

\emph{Cross-Site Scripting} (XSS) is a notorious class of vulnerabilities, consistently earning a place in the OWASP Top 10 as one of the most critical web security risks~\cite{OWASPTop10-2013, OWASPTop10-2017, OWASPTop10-2021}.
These vulnerabilities are rooted in the representation of web content, which only loosely separates data from markup and enables an attacker to manipulate a website or even execute code through malicious inputs deliberately breaking this separation. Therefore, special care has to be taken when inserting user-controlled data into web content.

This problem of XSS is further exacerbated by the wealth of representations available for describing web content. For example, HTML documents can be interspersed with resources spanning various languages, such as Cascading Style Sheets (CSS), JavaScript code, Scalable Vector Graphics (SVG) and even mathematical expressions (MathML). As a result, XSS vulnerabilities can arise in any of these formats if user-controlled data is inserted without sanitization.

However, this \emph{mishmash} of languages also imposes requirements on any attack exploiting XSS vulnerabilities. As each language defines its own syntactic and semantic rules, an attack payload has to be tailored to the exact location where it is inserted in the document. In the following, we refer to this location as the \emph{injection context}.
Similarly, protection mechanisms need to be context aware to correctly assess what constitutes dangerous input for a specific injection context.
This makes it challenging to defend against XSS~\cite{WeinSaxAkh2011,SaxMolLiv2011}.

To illustrate the role of the injection context, let us consider the running example in \Cref{fig:xss-contexts}. 
The code snippet contains three lines, each highlighting a different injection context: inside of an HTML \code{a} tag \dcircle{1}, as a URI in the \code{src} attribute in the \code{iframe} element \dcircle{2}, and as a double-quoted string inside a JavaScript code context \dcircle{3}.
Input injected at each of these three points is processed by a different parser, namely the HTML, the URI and the JavaScript parser, respectively.
Consequently, each of these injection contexts and their many variations require substantially different attack payloads tailored for the syntactic and semantic rules of said parser and context.

In order to successfully achieve code execution within one of these injection contexts, one needs to craft a payload that either directly calls a JavaScript function or moves into a parsing state where this is possible.
Especially the URI case highlights how not only the current parser state, but also the directly preceding ones are relevant for success.
For example, the \code{src} attribute of the \code{iframe} at \dcircle{2} is directly exploitable with a JavaScript URI, unlike the \code{src} of an \code{img} tag.
Instead, we would have to move to a different parsing state first by trying to break out of the surrounding quotes.

\begin{figure}[htb]
\begin{minted}[escapeinside=||,breaklines,frame=lines,framesep=2mm]{HTML}
<a href="..."> |\dcircle{1}| </a>
<iframe src|=\htmlsquote{\dcircle{2}}|></iframe>
<script>if (x == |\htmldquote{\dcircle{3}}|) { ... }</script>
\end{minted}

\caption{Code segments depicting injections in three different contexts: In HTML \dcircle{1}, in a URI \dcircle{2}, and in JavaScript code \dcircle{3}.}\label{fig:xss-contexts}
\end{figure}

Therefore, to craft an XSS payload an attacker has to deduce the injection context first.
This is usually done by either manual testing, where the tester analyzes the website's code, or automatic payload generation in the presence of perfect information.
The latter, in particular, received a lot of attention from researchers applying \emph{taint tracking} techniques to uncover these vulnerabilities~\cite{LekStoJoh13, StoPfiKaiLek+15, MelDasShaBau+18, SteRosJohSto+19, KleBarBen+22}.
In short, these approaches use a modified browser to mark all data entered via user-controlled \emph{sources} as \emph{tainted}.
This taint information is then propagated through the browser until it enters a \emph{sink} functionality, which potentially results in code execution.%

While the detection of \emph{reflected XSS} vulnerabilities has been thoroughly studied, the \emph{server-side stored XSS} variant, \ie, when the payload is not directly executed and instead first stored on the server, is still a hard-to-detect XSS type~\cite{eriksson2021black}.
The main challenge with this variant is that it requires reasoning about \emph{intra-page dependencies}: The page where the payload is inserted to be subsequently stored somewhere, might only be indirectly related to the page where the exploit actually triggers after it is read from storage.
However, as long as these two sides are publicly accessible, a manual tester or automated system can still incrementally adjust the stored attack payloads to explore the server-side processing and finally derive a working exploit.

One particularly challenging variation of stored XSS is the so-called \emph{\blindxss{}} (BXSS) vulnerability, where even this context information is no longer available. %

\subsection{The BXSS Attack Scenario}
\begin{figure}[tb]
\centering
\subfloat[]{\label{fig:xss:stored}\includegraphics[width=0.45\linewidth]{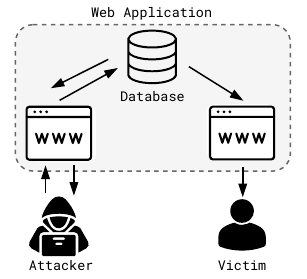}}
\qquad
\subfloat[]{\label{fig:xss:blind}\includegraphics[width=0.45\linewidth]{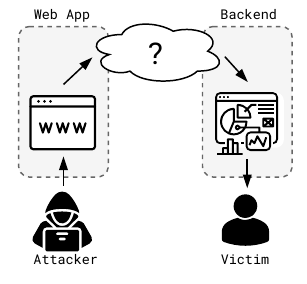}}\vspace{-1.5mm}
\caption{Attack flows of \emph{stored XSS} (a) and \emph{\blindxss{}} (b) vulnerabilities. Only the stored XSS scenario allows the attacker to inspect the injection context to adapt their exploit.}\label{fig:xss}
\end{figure}

As part of the deployment of larger web applications, there are often further web-based backends involved beyond what is accessible for visitors from the Internet.
Typical use cases for such intranet web applications are administration, monitoring, visitor statistics, and back office tasks.
While these applications are usually protected by a firewall from direct outside attacks, they can still be susceptible to XSS attacks if malicious inputs reach these internal sites.
For example, a monitoring system might read web server access logs from the public part of the website and display all \emph{Referer} headers on an internal webpage, allowing insights into visitor behavior.
However, if these attacker-controlled header values are not properly sanitized, they could result in XSS on that internal page.
Due to their purpose, the consequences of successful attacks on these internal pages can be especially high, as these pages are only visited by privileged users and might allow administrative actions.
Moreover, these backend systems might not be part of the application's threat model and therefore might have received less security attention.

The main challenge in this scenario from the attacker's perspective is that they have no information where their payloads might end up, hence the name \blindxss{}.
Without any feedback that allows them to identify the injection context in the backend's markup, dedicated XSS payload creation through means such as taint tracking is impossible.
Even worse, only a successful attack provides information on the vulnerability and allows to exfiltrate information about the backend. All unsuccessful attack attempts provide \emph{no} information to an attacker who is totally in dark about in which context the malicious payloads is injected, whether it has been adapted, or if simply nobody visited the backend system to trigger it.

\Cref{fig:xss} further illustrates this crucial difference:
As \Cref{fig:xss:stored} shows, the attacker can access the vulnerable page, learn about the injection context and adapt their payload accordingly in the stored XSS case.
However, they have no way of inspecting the admin panel in the backend in the \blindxss{} case depicted in \Cref{fig:xss:blind}.
Due to these challenges and to the best of our knowledge, \blindxss{} has not yet been systematically studied and no automated approach to uncover these vulnerabilities exists, so far.

\vspace{-1em}\paragraph{Threat model.} We consider the following threat model:
The attacker interacts with a publicly accessible website that might or might not be indirectly connected to an internal website via a storage mechanism, such as a database, log file or internal web service.
The attacker has no knowledge about any internal webpages, the used storage technology, or data transformations steps in between. 
There is no feedback unless the attack is successful.
The attacker's goal is to achieve code execution via XSS in the context of an internal web application of the website's backend, once visited by a person with access or an automated system with a full browser.
All other payload entry points besides publicly accessible websites, such as phishing emails, are out of scope for this work.

\subsection{XSS Polyglots}\label{sec:polyglots-background}
Ideally, we would overcome this challenge of not having a feedback by creating a \emph{universal XSS payload that triggers in all contexts}.
As previously described, we need to consider several different parsers for this, as our payload could end up inside any of them.
A construct that is valid input in multiple formats and/or languages is a so-called \emph{polyglot}~\cite{magazinius2013polyglots}.
For example, GIFAR is a technique to create a file that is both a valid GIF image and a valid Java Archive (JAR)~\cite{brandis2009exploring}.

In the context of this paper, \emph{polyglot} refers to an XSS exploit that is able to execute in multiple injection contexts, such as HTML and JavaScript code.
Moreover, \emph{payload} refers to the part of the exploit that is actually executed, \eg, \code{alert()} or \code{import()}.
Coming back to our example in \Cref{fig:xss-contexts} and assuming we want to execute the payload \code{alert()} for demonstration, the three different injection contexts require different conditions for the payload to be executed.

For the HTML context inside the anchor tag in injection \dcircle{1}, we first need to close that tag and then need to insert an element that executes code, \eg, like this: \code{</a><script>alert()</script>}.\footnote{The dangling \code{</a>} will be ignored by the browser.}
On the other hand, the second injection context expects a URL \dcircle{2}.
While we could try to break out of the surrounding quotation marks, this might not be possible due to sanitization.
One solution is to instead use a JavaScript URI, \eg, by supplying \code{javascript:alert()}.
Finally, the injection context in JS code \dcircle{3} requires us to break out of the quotes and close all expected parentheses to make sure that the code does not break, \eg, \code{")\{\}alert();//}.
The two trailing slashes are important to comment out the rest of the line to avoid a syntax error.

For this simple example, we can still come up with a polyglot by carefully combining all three exploits by hand: \code{javascript:alert()//")\allowbreak\{\}alert();//\allowbreak</a>\allowbreak<script>alert()</script>}. %
For case \dcircle{1}, this just puts a lot of ``garbage text'' into the anchor text before closing the tag and executing the script.
For case \dcircle{2}, this is interpreted as a URI with valid JavaScript code and a long trailing comment.
Finally, for injection \dcircle{3}, this puts the URI in the string, closes the \code{if}-branch and executes our payload, while ignoring the rest due to the second \code{//} comment.

However, this polyglot just covers three contexts and there are a lot of variations where it would not work properly.
For example, injection \dcircle{1} could also happen in the \code{href} attribute, with either single or double quotes, or might be inserted via \code{innerHTML} where the script tags do not execute but event handlers do~\cite{whatwg-html-parsing-flags}.
As another example, \dcircle{3} would result in a syntax error when the code has an \code{else} branch, as our inserted \code{alert} would break the \code{if-else} construct.
Covering \emph{all these common variations of all contexts} results in an explosion of possible combinations, making manual polyglot construction extremely tedious, if not impossible.

\begin{figure}[htb]
\begin{minted}[escapeinside=||,breaklines,frame=lines,framesep=2mm]{JavaScript}
//Read from a stored object without injection into the source
document.location = someObj.redirectLocation;
//Directly injected into the source code of a .js file
var foo = |\dcircle{4}|;
\end{minted}

\caption{Two injection contexts that are mutually exclusive, \ie, can not be solved by the same XSS polyglot.}\label{fig:mutually-exclusive}
\end{figure}

As an additional complication, some injection contexts are \emph{mutually exclusive}.
\Cref{fig:mutually-exclusive} shows one example of such a case: The first injection context is only possible to solve by exploits starting with \code{javascript:} as such a JavaScript URI does not trigger an actual navigation, but instead executes the subsequent code in the context of the current document.
Starting with anything else will either fail, as escaping from the statement is impossible without a direct reflection into the source code, or cause a navigation away from the website that was the target of the attack.
On the other hand, there are also injection contexts that do not allow \code{javascript:} at the very beginning, \eg, the injection context in \Cref{fig:mutually-exclusive} at position \dcircle{4}, as the colon produces a syntax error on the right side of the variable assignment.
Illustrated by these two examples, mutually exclusive injection contexts thwart efforts to construct \emph{one} universal XSS polyglot.
On the other hand, full coverage could be achieved with one payload per context.
However, we aim at creating a small \emph{set of polyglots} while still covering the entire state-of-the-art testbed.
From a tester's perspective, this results in less required tests and less system log noise.
From a researcher's view, it means less submissions and thus less noise and reduced load for systems-under-test.

\subsection{Research Questions}\label{sec:rqs}

Constructing a set of powerful polyglots is key to the analysis of \blindxss{}, as it makes it possible to successfully trigger vulnerabilities without feedback. While previous work has focused on manually engineering polyglots \cite{ultimate-polyglot, szurek-polyglot, ostorlab-polyglot}, we argue that an automated process is inevitable here to handle the amount of injection contexts. This leads us to the following research questions that structure our work:
\begin{questions}
\setlength\itemsep{-0.3em} %
	\item Is it possible to automatically create a set of XSS polyglots that are capable of covering all common injection contexts in combination?\label{rq:1}
	\item Do these synthesized polyglots scale up to real-world settings and how do they compare with existing XSS detection approaches?\label{rq:2}
	\item Can these polyglots be employed to perform a large-scale analysis of \blindxss{} and indicate vulnerabilities in web application backends?\label{rq:3}
\end{questions}

To address these research questions, we conduct a series of experiments:
In \Cref{sec:generation}, we introduce a method for synthesizing polyglots using concepts from game theory~(\ref{rq:1}).
In \Cref{sec:taint-validation}, we assess the real-world performance of these polyglots, comparing them to traditional approaches~(\ref{rq:2}).
Finally, in \Cref{sec:bxss-study}, we perform a large-scale study to detect BXSS vulnerabilities in the wild~(\ref{rq:3}).

\section{Synthesizing Polyglots}\label{sec:generation}
The generation of a polyglot is a challenging task. 
Unlike attack payloads designed for a single environment, we are faced with multiple programming languages and contexts.
Although a polyglot is basically just a string of characters, conceptually it is a chimera made up of terminal symbols from different grammars. Some of these characters are even ambiguous and only parsed correctly when the respective injection context is known.
To cope with this complexity and ambiguity, we develop an automated approach that phrases the synthesis of a polyglot as a discrete optimization problem.
The objective is to find a string that maximizes the number of exploitable injection contexts on a given testbed, regardless of the underlying languages and grammars.

This optimization has several constraints:
First, evaluating a polyglot on the testbed involves running an entire browser, including HTML parser and JavaScript engine.
Second, during the polyglot evaluation, we receive only binary feedback indicating success or failure, making gradient-based optimization unfeasible.
Lastly, as discussed in \Cref{sec:polyglots-background}, some test cases are mutually exclusive. Therefore, we must identify a \emph{set} of polyglots that collectively solve the testbed.

\subsection{Monte Carlo Tree Search}%
\label{sec:mcts}%

Interestingly, the described optimization problem can be cast into a game in which a player iteratively modifies a polyglot to exploit as many contexts as possible. In this game, moves correspond to changes of the polyglot, while its reward is determined by the number of exploited contexts. Different algorithms exist for solving such games, ranging from simple search strategies to reinforcement learning \cite{russell2010artificial}.

For our analysis, we focus on the concept of \montecarlo{} (\mcts{})~\cite[][]{browneMcts:2012} that is known to find strong moves in round-based games, such as Chess and Go \cite{silver2016mastering}. The generation of polyglots, however, is agnostic to this algorithm and thus we present a comparison of \mcts{} with other strategies for solving the underlying game in Appendix \ref{appendix:alternative-generation}. %

Technically, \mcts{} simulates multiple games to conclusion to gather knowledge about promising game states.
To keep track of the performance, \mcts{} constructs a game tree in which each node corresponds to a polyglot and contains two attributes: a visit and a win counter. The algorithm explores a path in this tree using four steps:

\begin{enumerate}
\setlength\itemsep{-0.3em} %
\item \emph{Selection:} Starting from the root node of the game tree, an unexplored child or the most-promising child node is selected until a leaf node is reached.
\item \emph{Expansion:} If the selected leaf node is non-terminal, the game tree is expanded by generating the child nodes of the leaf node.
\item \emph{Playout:} Starting from the leaf node, random nodes are explored until a terminal node is reached. This is equivalent to performing random moves until the game ends.
\item \emph{Backpropagation:} On the path back to the root, the visit counter is incremented by one and the win counter is updated according to the result of the simulated game.
\end{enumerate}

During the selection phase, \mcts{} balances exploitation (selecting moves that were successful in the past) and exploration (gathering knowledge about unexplored areas).
To rank each child $i$, we use the upper confidence bound $\frac{w_i}{n_i} + \sqrt{\frac{2 \ln N_i}{n_i}}$ where $w_i$ is the number of wins, $n_i$ is the number of visits and $N_i$ is the number of visits of the parent node. If a child has not been explored before, \ie, $n_i=0$, it takes precedence.

\definecolor{gfrcolor}{HTML}{7DBA91}
\definecolor{gfrred}{HTML}{8D0400}
\definecolor{gfryellow}{HTML}{F8CC00}
\definecolor{gfrdiamond}{HTML}{297A8C}

\begin{figure*}[ht]
	\centering
    \includegraphics[width=\linewidth]{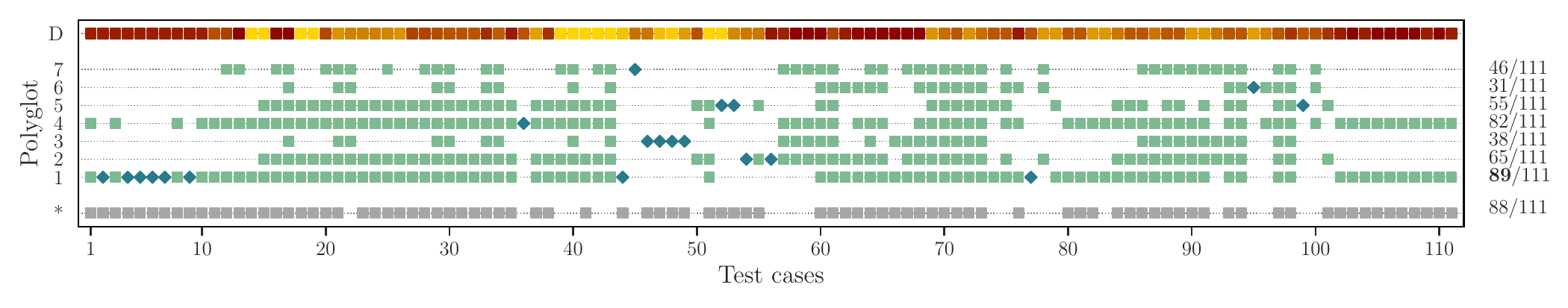}\vspace{-3mm}%
	\caption{%
	Composition of our minimal polyglot set solving all \num{111} test cases of our testbed. 
	Squares (\textcolor{gfrcolor}{$\blacksquare$}) indicates solved tests.
	The top row (\textrm{D}) visualizes the overall difficulty of each test case, calculated from the total ratio of polyglots we generated that solve this test \ie, \emph{difficult tests} with fewer solutions are red (\textcolor{gfrred}{$\blacksquare$}), while the color of more \emph{frequently solved} tests gradually shifts to yellow (\textcolor{gfryellow}{$\blacksquare$}).
	The numbered rows show the performance of our polyglot set.
	Diamonds (\textcolor{gfrdiamond}{$\blacklozenge$}) are placed instead of squares, where only one polyglot in the set solves a test.
	For comparison, the bottom row (*) shows the performance of the Ultimate polyglot. 
	}%
    \label{fig:set-gfr-perfomance}
\end{figure*}

\subsection{Synthesizing Polyglots with \mcts{}}%
\label{sec:synthesizing-with-mcts}

To reduce the extensive search space, we incorporate expert knowledge into the construction of the game tree:
Instead of operating on all characters, we generate polyglots only from a set of tokens, such as \code{<} and \code{script}. We refine these tokens with simple rules that prevent invalid combinations form appearing, such as \code{svg} appended to \code{iframe}.
As a result of this reduction, \mcts{} omits moves that are impossible to yield reasonable polyglots.
The resulting token set consists of HTML literals, HTML tag names, HTML event handler content attribute, and other HTML tokens.
The complete list including the rules can be found in our companion repository. %

Starting from an empty string, \mcts{} iteratively appends tokens to it to expand the game tree and measure the polyglot's reward.
Unlike conventional games, however, this game has no clear end state, since polyglots can be arbitrary long.
We therefore set a fixed limit of 400 characters, after which we stop appending tokens to the polyglot.
Since this fixed length condition might lead to superfluous tokens, we employ a minimization strategy afterwards (see Section~\ref{sec:polyglot-minimization}).

To evaluate a polyglot during search, we implement a fast, manually created testbed covering \num{27} common XSS injection contexts using Puppeteer (13.1.3) based on Chromium (98.0.4758.0).
The complete testbed is included in our companion repository.
In the regular \mcts{} backpropagation, each polyglot would be assigned a score used to update the win counters on its path. Since we evaluate the polyglot on multiple test cases simultaneously, however, we receive a list of scores on each playout. Instead of saving only one score, we thus save the entire list and sum it up, once we need the total score $w_i$ for a particular node in the tree. As a result, we can keep track of the individual tests solved by a particular polyglot during construction.
This gives us more flexibility: When some test cases have already been covered, we can exclude them from the score calculation in later runs. %

The final synthesis of one polyglot is shown in Algorithm~\ref{alg:gen-polyglots}:
Starting from the root node, we use \mcts{} to explore the game tree for $N$ rounds until we fix the first move, \ie, select the first token of the polyglot.
With the first token in place, we repeat the process $N$ times and select the second token.
We continuously start the game from a deeper root node until the depth limit $D$ is reached.
At the end, the polyglot with the highest score is returned.

\begin{algorithm}[htb]
\caption{Generating a polyglot}%
\label{alg:gen-polyglots}
\tablesize
\DontPrintSemicolon
\SetKwInput{KwInput}{Input}
\SetKwInput{KwOutput}{Output}
\SetAlgoNlRelativeSize{-2} %
\SetNlSty{normalfont}{}{} %
\KwInput{root, start node for generation}
\KwInput{$D$, maximum depth of the root node}
\KwInput{$N$, iterations until a move is chosen}
\KwOutput{polyglot string}
bestPolyglot, bestScore $\gets$ null, 0 \;
\For{$i \gets 1$ \KwTo $D$}{
    \For{$j \gets 1$ \KwTo $N$}{
        leaf $\gets$ select(root)\;
        expand(leaf)\;
        polyglot, score $\gets$ playout(leaf)\;
        backpropagate(leaf, score)\;

        \If{score > bestScore} {
            bestScore, bestPolyglot $\gets$ score, polyglot \;
        }
    }
    root $\gets$ choose\_best(root)\;
}
\KwRet~bestPolyglot\;
\end{algorithm}

At best, our approach would generate a single polyglot to ``rule them all'', that is, solve all provided test cases.
However, our analysis reveals that depending on the test cases, this might not be possible.
For example, as discussed in \Cref{sec:polyglots-background}, some injection contexts are mutually exclusive and a polyglot providing proper execution of JavaScript code in both cannot exist.
As a remedy, we develop a strategy for combining multiple polyglots into a set. 

In particular, after one polyglot is created, we remove the test cases it solves and start over the search process of \mcts{}.
As result, by design, we seek a complementing polyglot that focuses only on those cases the previous one cannot solve.
We continue generating new polyglots using Algorithm~\ref{alg:gen-polyglots}, until all tests have been covered or a maximum number of tries is reached.
Finally, we obtain a set of polyglots that solves the entire testbed of the synthesis process.

\subsection{Selecting a Final Polyglot Set}%
\label{sec:polyglot-evaluation}

While we synthesize polyglots on a fast testbed, we use a larger testbed to determine the quality of the generated polyglots.
For this purpose, we set up a local instance of the Google Firing Range (GFR)~\cite{github-firing-range}, which is a state-of-the-art XSS testbed~\cite{BazCriMag16}.
We manually choose a subset of GFR tests for our scope, based on the following criteria:
First, we exclude non XSS-related test cases.
Second, in cooperation with Google, we obtain a list of GFR tests that have a known solution which we manually confirm.
We further exclude tests exceeding our scope, namely exploits relying on specific frameworks and those requiring a specific encoding.
Tests that aggressively block special characters with an error page, \eg, any request that contains \code{<} or \code{>}, are also excluded.
While solutions for these tests exist, their solution needs to be so narrow that they are no longer a polyglot and thus also out-of-scope. 
We confirmed all solutions and verified that an applicable solution exists for our requirements.
A detailed list of the excluded tests and the specific reasons can be found in Appendix~\ref{appendix:gfr-exclusions}. %

After generating \num{4000} polyglots over multiple runs from different seeds, we utilize a greedy min-set cover approach to select a minimal subset that solves all GFR test cases.
We build this minimal set by adding the polyglot to the set that solves the most tests yet unsolved by the set, until no additional solutions can be added.
Our final set of polyglots that is used in our study consists of 7 instances.
Thus, we answer \ref{rq:1} and show that it is possible to automatically create a set of XSS polyglots that cover common injection contexts.
To compare the efficacy of our final polyglot set with a publicly available polyglot, we chose the best polyglot from the blog posts mentioned in \Cref{sec:rqs}---the \emph{Ultimate \xss{} Polyglot}~\cite{ultimate-polyglot}---as our reference. 
Now, to display the composition of our polyglot set, \Cref{fig:set-gfr-perfomance} visualizes our evaluation results for each polyglot on the \num{111} test cases.
We give an indication of the difficulty of synthesizing a solution for each test by assigning a color to each of them. 
The number of polyglots from our full set that successfully solve a test, determines its difficulty.
Tests for which we synthesized fewer solutions tend to be more challenging and are depicted in red.
In contrast, the color of easier tests, \ie, those with more solutions, gradually shifts to yellow.
The numbered rows below correspond to each polyglot of our final set, while the bottom row displays the Ultimate polyglot's score.
Indicated by diamond symbols in the corresponding rows, each polyglot in the final set contributes at least one unique solution, \eg, polyglot \num{4} is the only in the set solving test \num{36}.
The plot shows that our polyglots cover all test cases, ranging from \num{31} to \num{89} tests covered by each individual polyglot.
The Ultimate polyglot works surprisingly well and manages to cover \num{88} tests.
However, our automatic polyglot set creation outperforms manual engineering, as our generation process can target gaps left by previously synthesized polyglots.

\subsection{Minimizing Polyglots}%
\label{sec:polyglot-minimization}

Since our generation uses a termination criterion that sets a fixed minimum length for a polyglot, it is possible that some tokens in the polyglot are superfluous.
In general, this would not constitute a problem, but some websites have length constraints to the inputs in their backend. %
To make sure that such obstacles do not obstruct our polyglots, we minimize the lengths of our final polyglots.
The objective is to find the smallest polyglot with equivalent test results on the GFR testbed\@.
The minimization is done automatically and at token level, \ie, we first deconstruct each polyglot into its tokens.
Since each token could be removed independently, there are $2^N$ minimization candidates where $N$ is the number of tokens in the polyglot.
Instead of testing all $2^N$ options, we build our minimization strategy on a single assumption:
If removing a token changes the test results, then all minimizations in which that token is removed are invalid.
Although there might be edge cases, where this does not hold true, this drastically reduces the search space as we can first test each token separately.
Our polyglot minimization then involves iteratively removing combinations of tokens until no more tokens can be removed without altering the evaluation result.
Overall, in every polyglot (except for one), some unnecessary tokens have been identified and removed.
For the other polyglots we achieve a reduction between 6\% and 24\%.

\section{Validation on Real-World Websites}\label{sec:taint-validation}
While our test bed and the Google Firing Range cover a wide range of \xss{} vectors, they can only serve as an artificial evaluation that is not necessarily representative of the polyglots' ability to discover vulnerabilities in real-world code.
Thus, in this chapter we set out to validate how well our \xss{} payloads perform on actual websites with non-trivial codebases and to compare them to state-of-the-art techniques for detecting XSS that leverage targeted exploit generation.

More precisely, we evaluate our \xss{} polyglots on a set of recently found real-world client-side \xss{} (CXSS) vulnerabilities. 
CXSS has several advantages in the context of our experiments: 
While exposing the same characteristics as other classes of \xss{}, especially in respect to injection points and syntactical restrictions, it allows precise exploit payload generation, thanks to the availability of full data-flow information~\cite{LekStoJoh13,StoPfiKaiLek+15,MelDasShaBau+18} (see Sec.~\ref{sec:exploit-gen} for details). Furthermore, research experiments utilizing CXSS offer the invaluable benefit that the complete exploitation attempt is conducted solely on the client-side. Hence, potential negative side effects on the real world websites can reliably be prevented.

\subsection{Targeted \xss{} Exploit Generation}\label{sec:exploit-gen}
As the baseline for the prevalence of CXSS vulnerabilities, we re-use a state-of-the-art taint tracking engine~\cite{foxhound} and the associated exploit generation by \citet{BenKleBarJoh21}.
In the following, we give a short overview of their approach, using the code in Figure~\ref{lst:xss-example} as an illustrative example throughout.

\begin{figure}[htb]
\begin{minted}{javascript}
let d = document.getElementById("..");
let href = decodeURIComponent(window.location.href);
d.innerHTML = '<a href="' + href + '">';
\end{minted}

\caption{A typical Client-Side \xss{} vulnerability.}\label{lst:xss-example}
\end{figure}

In general, the idea is to browse the web with a modified taint browser while constantly monitoring and analyzing all data flows.
Once the taint browser detects a data flow that is potentially susceptible to CXSS it has the following information available:
The source (here \texttt{location.href}), the sink (here \texttt{.innerHTML}), what characters from the source ended up in the sink, as well as the whole string entering the sink.

With this information, the exploit generation strategy works as follows:
The first step is to generate the so called \emph{breakOut} sequence, which aims to close the current context (here the double quoted \code{href} attribute) and put the parser into a state where we can insert a \xss{} payload.
Several HTML tags, such as \code{iframe} or \code{textarea} prevent code execution of their children. So the \emph{breakOut} will close those tags as well.
Next, a context specific payload is generated, this is based on the sink function, as \code{innerHTML} requires a different payload than, \eg, \code{eval()}. The payload generated by the exploit generator for \code{innerHTML} is \code{<img src=x onerror=f()>}.
If this report function is called, we know that the generated exploit was successful.
However, with remaining markup from the original \code{<a>} tag, we must first create  a \emph{breakIn} sequence.
Its purpose is to ``consume'' the leftover characters , ensuring the parser proceeds without errors.
While this step is usually unnecessary in the HTML context due to the parser's leniency, it becomes critical in the JavaScript context.
A suitable \emph{breakIn} would be to comment out the remainder of the document.

Finally, these three parts (\emph{breakOut}, generated payload and \emph{breakIn}) are concatenated and inserted into the URL at the appropriate position.
For our running example, the insertion point would be to simply append the generated exploit in the fragment of the URL.\@
Therefore, and in contrast to our polyglots, each exploit generated with this approach is designed to work for \emph{one} specific data flow on \emph{one} website only.

\subsection{Validation Experiment Setup}\label{sec:tainting-experiment}
To compare the performance of different approaches on real-world websites, we surveyed the top \num{10000} domains according to the Tranco list~\cite{le-pochat-tranco} as of Dec. 15, 2022, available at \url{https://tranco-list.eu/list/W95V9}.
For each successfully visited site, we queued up to \num{10} subpages, enabling us to capture vulnerable data flows that might not be apparent on the landing page.
All relevant data flows for CXSS were stored in a database, and we concurrently visited the generated exploit URLs to validate them.
The exploit validation is done with legacy URL encoding, consistent with previous works~\cite{LekStoJoh13,MelDasShaBau+18,SteRosJohSto+19,KleBarBen+22}.
For each URL generated by the exploit generator, we systematically replaced the targeted exploit with each of our 7 synthesized polyglots.
Subsequently, we used our crawler to visit these URLs, flagging the exploit as successful if the callback function was triggered.
To assess the effectiveness of our synthesized polyglots in comparison to publicly available ones, we subjected the Ultimate \xss{} Polyglot to the same treatment, like in \Cref{sec:polyglot-evaluation}.
The Ultimate polyglot, like our polyglots, operated without the additional information provided by the taint tracking browser, distinguishing it from precisely generated tainting exploits.

\subsection{Comparison of XSS Detection Rates}\label{sec:tainting-evaluation}
\begin{figure}
    \centering
    
    \includegraphics[height=4.25cm,keepaspectratio]{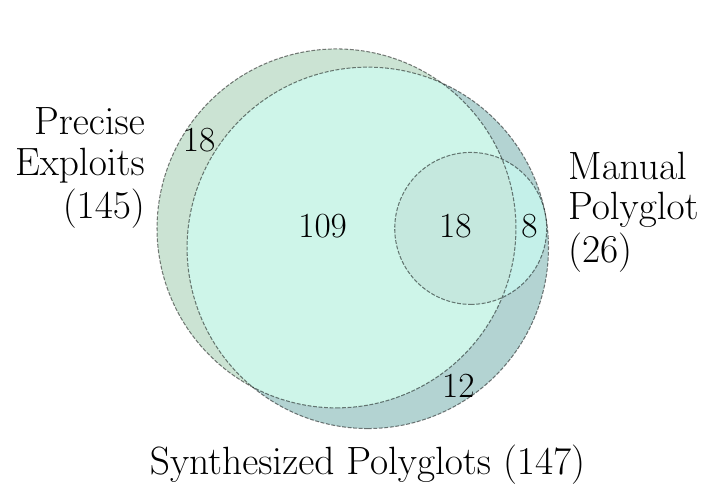}\vspace{-1.5mm}%
    \caption{Euler diagram of the performance of the three different XSS detection approaches on the Tranco Top-\num{10000}.}%
    \label{fig:client-side-xss-venn}
\end{figure}

In total, we generated exploit URLs for \num{1010} of the visited websites due to potentially security sensitive data flows.
We then applied each of the three approaches to see if they can achieve JavaScript execution.
Thereby, we were able to successfully validate vulnerable data flows on a total of \num{165} websites, resulting in XSS\@.
Figure~\ref{fig:client-side-xss-venn} depicts the results of how our synthesized polyglots perform compared to both the perfect-knowledge exploit generation strategy, as well as the ultimate polyglot from prior work, thus answering~\ref{rq:2}.
As the figure shows, on \num{127} out of the \num{165} websites both our polyglots and the exploit generation were successful.
At the same time, it also highlights that both approaches have their advantages as not a single one was able to discover all exploits.
However, the ultimate polyglot had the worst performance by far, as it only worked on a fraction of the websites overall while not finding any additional exploits that were not already covered by the other approaches.

We then proceeded to manually investigated the successful exploits exclusively solved by either of the approaches.
Examples of data flows where the polyglots did not work as well as the generated exploits were highly specific edge cases such as the one shown in Figure~\ref{lst:tainting-gen-sample}, the root cause behind 83\% of the websites where the polyglots were unsuccessful but the precise exploit generation succeeded. 
While this looks like a simple \xss{} vulnerability that the polyglots should be able to solve, it is important to note that the dynamically created \texttt{div} is never added to the DOM\@. 
This prevents triggers that rely on the \texttt{onload} event handler to fire.
Another reason for the polyglots failing to trigger results from the general concept of a polyglot (see Section~\ref{sec:polyglots-background}).
To be valid code in several contexts, they contain syntactic elements of several programming languages.
Application code trying to parse the tainted data can break, causing the vulnerable code paths to never be executed at all.
Take the polyglot from \Cref{sec:polyglots-background} as an example.
If \mintinline[]{javascript}{data.split(':')[0]} is called on it, it extracts everything in front of the double colon, e.g., to retrieve a field from a complex data structure.
For said polyglot this returns \code{javascript}, a harmless string.
The precise exploit generation never adds non HTML characters such as double colons, and as such its payload would be included in the extracted text. %
However, their rich syntactic structure is a positive in other cases, because they are able to evade broken sanitizing routines.
One common mistake for input sanitization is misunderstanding the \texttt{replace} semantics and only replacing the first occurrence of the needle by mistake according to \citet{KleBarBen+22}.
Due to containing several code execution triggers, our polyglots are able to evade such a faulty sanitization routine whereas the generated exploit does not.

\begin{figure}[htb]
  \begin{minted}{javascript}
function read_href(url) {
  var div = window.document.createElement('div');
  div.innerHTML = '<a href="' + url + '"></a>';
  return div.firstChild.href;
}
read_href(window.location.href); // somewhere else
\end{minted}
\caption{Problematic data flow not solved by our polyglots.}%
\label{lst:tainting-gen-sample}
\end{figure}

After demonstrating our polyglots' exceptional performance in detecting CXSS flaws on real websites, we investigate their suitability to discover \blindxss{} in the next section.

\section{Blind XSS in the Wild}\label{sec:bxss-study}

So far, we have demonstrated how to synthesize polyglots and have found that their effectiveness in identifying reflected XSS is comparable to existing approaches.
In this section, we design and conduct the first large-scale study of \blindxss{} in the wild.
Our study is based on the synthesized polyglots.
Hence, it demonstrates their unique advantage in uncovering BXSS vulnerabilities.
Additionally, we show the importance of using a set of polyglots instead of a single one.

We are aware of our responsibility to conduct this study in an ethical manner and, in accordance with the Menlo report~\cite{KenDit12}, we designed it to prevent harm.
Most importantly, if a polyglot should trigger, it will only connect back to inform us about its execution.
These benign tests allow us to measure the scope of the problem and warn website owners to close security vulnerabilities before they can be exploited.
We consulted our funding project's institutional review board (IRB) which concluded that the potential gain for system security predominates potential extra work for website operators.
Further technical details are found in \Cref{sec:ethical}. %

\subsection{Polyglot Preparation and Monitoring}\label{sec:monitoring-notification}
A polyglot triggers all vulnerabilities it can cover.
However, due to the nature of \blindxss{} we cannot directly observe when this happens in a backend.
Hence, we design the polyglots to notify our monitoring server if they are executed due to a vulnerability.
We outsource this notification mechanism into a remote \emph{\ourscriptname{}} instead of embedding the required functionality into the polyglots for multiple reasons.
To begin with, using a replaceable remote script makes polyglot synthesis easier, as we only have to optimize on import mechanisms as our polyglots' core functionality.
Additionally, testing is sped up, as the imported code can be easily replaced without syntactically changing the polyglots. 
In contrast, embedding the required notification functionality would introduce additional characters to the polyglots which would increase their length and extend the required character set which in turn may trigger additional input filters potentially reducing the polyglots effectiveness in the field.
Finally, using a remote script enables us to render polyglots ineffective by stopping to serve the \ourscriptname{}, e.g., once the study concludes after a certain period. %

To achieve traceability from polyglot submission to individual vulnerabilities we assign each submission a unique \num{12}-character ID\@.
This ID encodes how (URL, form, or header) which polyglot was submitted to and on what (backend) page it ended up being executed.
This ID is embedded in the URL a polyglot's import functionality takes, e.g., JavaScript \mintinline[]{javascript}{import} statements or the \mintinline[]{javascript}{src} attributes.
Exemplarily, a polyglot requests the notification script from our monitoring server upon execution, \eg, via \mintinline[]{javascript}{import('https://<ID>.<monitor_host>/s.js')}.
The monitoring server can then extract the ID from the requested script's URL and embed it into the notification script before returning it.
The script in turn includes this ID in the feedback ping it returns back to our monitoring server when executed.
Using this submission ID, a backend vulnerability can be linked to a specific submission on a particular website.

\vspace{-1em}\paragraph{\Ourscriptname{}} The \ourscriptname{} (\Cref{appendix:probing}) returns information required for accurate detection and effective disclosure of a \blindxss{} vulnerability.
When executed, the script returns the document's title, its URL, excluding query and fragment, as well as the JavaScript user agent and platform.
It encodes the information, the submission ID, and the current timestamp in the URL of an HTTP request bound to our monitoring server.
Upon receipt of such a request, we indirectly receive the IP address of the sender, which we require for the disclosure process. 
\Cref{sec:ethical} further discusses the usage and implications of the collected data. %

\pagebreak
\subsection{Polyglot Transmission}%
\label{sec:polyglot-transmission}

There are three general ways to transmit our synthesized polyglots to the websites we visit, where they can get passed to backend systems: \emph{headers}, \emph{URLs}, and \emph{forms}.
In the following, we will briefly outline considerations for each of them.

\vspace{-1em}\paragraph{HTTP Headers.} 
To test backend logging applications, we submit polyglots in four request headers. 
The {Referer} header, often containing the previously visited URL, is valuable for analytics and tracking~\cite{mdn-http-headers-referer}.
Similarly, the {User-Agent} header aids analytics, by revealing browser and platform usage patterns.
Polyglots in the {Cookie} header may get logged for failed authentication.
Additionally, we utilize the less-known {Warning} header, particularly warning code \num{199}, used to transmit loggable information~\cite{mdn-http-headers-warning}.
Despite its deprecation, all major browsers still support it.
We employ HTTP GET requests to transmit each header, embedding the polyglots as either direct values or in suitable contexts like \code{Cookie: test=<polyglot>}.
To identify which header triggers feedback, we individually send each polyglot ro the landing pages via a GET request for each header mentioned.

\vspace{-1em}\paragraph{HTTP URLs.} 
Similarly to headers, URL submissions via the \emph{query} and \emph{path} are issued for each website with each polyglot.
For every landing page visit, we append an artificial subpath, followed by the polyglot as another subpath, \eg, \code{http(s)://domain.com/\ul{<path>}/\ul{<polyglot>}}.
Analogically, in query submission, we employ an artificial query key with the polyglot as its value, \eg, \code{http(s)://domain.com?\ul{<query>}=\ul{<polyglot>}}.
We use artificial paths and query keys, because our goal is not to find reflected and stored XSS flaws in existing functionality.
Instead, our aim is to transmit our payloads to web-based backend systems with potential \blindxss{} vulnerabilities.
Moreover, we perform this action only once on the landing page, avoiding excessive warnings for website operators.

\vspace{-1em}\paragraph{HTML Forms.} 
Finally, we also analyze the HTML code for each page that we visit and extract all contained HTML forms.
For each form, we first check if the allowed length of each input given by the \mintinline[]{javascript}{maxlength} attribute is enough for our longest polyglot.
Moreover, we have measures in place to prevent duplicate form submissions.
On par with previous work~\cite{ssb-2022}, a form is considered new if at least one value differs from previous forms: (a) its \mintinline[]{javascript}{innerHTML} representation (excluding default values and whitespace), or (b) the form's target domain.
For all the remaining unique forms, we fill all inputs with one polyglot at a time and submit the form.

\subsection{Identifying Blind XSS}\label{sec:bxss-filter}

\noindent
Generally, we expect a mix of automated business logic tools, and manually operated monitoring and administration platforms to trigger our polyglots.
The former may react to our submissions instantly, while the latter may be bound to human interaction and thus only trigger sporadically.
Therefore, we give each submission a time frame of \num{2}~months during which we monitor it.
In the following, we explain our approach to confirm that triggered polyglots are of the \blindxss{} type.

We define three cascading filter steps to narrow down our findings to \blindxss{} and thus discarding reflected and stored XSS on the way:
(1) Feedback pings have to come from an IP address different to our crawler's IP, otherwise we triggered a reflected XSS\@.
(2) The URL where we submitted our polyglot has to differ from the URL where it triggers, otherwise we found a trivial stored XSS where the attacker could have adjusted their payload in this non-blind setting.
(3) The URL where the polyglot triggers may not be publicly accessible, only then we have discovered a \blindxss{} vulnerability with no way for the attacker to learn about the injection context.
Otherwise, we found a non-trivial stored XSS, where the two URLs differ and their connection needs to be discovered first, but both are nevertheless publicly available.

To test this, our monitoring server initiates an additional confirmation step for each newly reported BXSS candidate URI immediately after receiving a feedback ping.
In this step, the server conducts an extra visit to the reported URI to assess its public accessibility.
At this stage, our filter (3) confirms invalid or local URIs, as well as private IP addresses as BXSS instances where our polyglot reached and executed in a backend system.
However, even if the page did load, some cases may still qualify as BXSS\@.
For instance, we encountered public pages that required authentication to access their content.
Since the identification of login pages is hard to automate~\cite{jonkerShepherdGenericApproach2020, drakonakis2020cookie}, we manually investigated and labeled these websites as either BXSS or false positives.

\subsection{Ethical Considerations}\label{sec:ethical}
Conducting server-side studies requires careful ethical consideration.
To this end, we followed best practices outlined in the Menlo report~\cite{KenDit12}, and aimed at uncovering real-world BXSS scenarios while ensuring minimal impact on operators and users.
We decided on a large-scale study without the operators' consent. %
This is a difficult and controversial decision that requires a thorough investigation of the potential harms and risks.
We discussed this decision in detail with our IRB and received approval.
Nevertheless, we recognize the weight of this decision and the potential for alternative options, the advantages and disadvantages of which we discuss below.

\vspace{-1em}\paragraph{Alternative study designs.} %
Auditing \emph{open-source applications} for BXSS vulnerabilities offers a first choice. This method comes with no ethical issues but its results are limited by the proprietary nature of production code and the unpredictable configurations of live websites.

Analyzing simulated backends in a \emph{lab environment} offers another choice.
While ethically unproblematic within a confined environment, accurately mirroring the nuances of real-world backends poses challenges, often necessitating insights from active operators on their particular setups.

Real-world studies, especially a \emph{small-scale study} with prior operator consent, emerge as another solution, allowing operators to take precautionary measures to minimize harm.
Given the---at the time of our study---unknown prevalence of BXSS and indications of its rarity~\cite{eriksson2021black,doupe2010johnny,bau2010state,parvez2015analysis}, we decided to discard this study design.
However, an appropriate selection of operators might provide representative results.

We chose the final option, conducting a \emph{large-scale study} of the top-ranked website without acquiring operator consent.
While offering a direct and unbiased approach to a representative analysis of the subject, it is important to note that this strategy is ethically problematic, even if tests are extremely carefully designed. 
We remark that the other alternatives would also be applicable with different compromises between ethical implications and gained insights.
In general, we recommend thoroughly considering different study designs and additionally engaging with an IRB upfront. 
Thus far, our study has incurred no reported damages or problems, attesting to the effectiveness of our design in preventing harm.
Our results revealed \num{20} vulnerabilities among the top \num{100000} websites underlining our initial assumption of the rarity of BXSS\@.
Nonetheless, it is imperative to stress that any retrospective analysis does clearly not provide justification for the decisions made in a study.

It is crucial to highlight that conducting studies without proper consent or assuming consent from non-responsive parties is not ethically sound. Our research should thus \emph{not} be viewed as a template for similar studies. Based on our discussions with the IRB, we believe that our work represents a good compromise in this regard. However, we acknowledge that alternative study designs could have been employed to mitigate the risk of harm more effectively, though at a higher risk of reduced insights. Scientific work often navigates complex terrains, demanding thoughtful balancing of conflicting interests. We selected---to the best of our knowledge and belief---a suitable balance, which is certainly not without debate.

\vspace{-1em}\paragraph{Side effects.}
All test requests contain minimal JavaScript payloads that do not affect the global namespace of the surrounding application ensuring that no unintended side effects on legitimate code occur.
In the rare cases where the initial polyglot succeeds, a second script is retrieved from our servers for data collection.
This two-step process enables us to deactivate the \ourscriptname{} for specific IDs or entirely at any time, serving as an additional \textit{mitigation strategy}. %

As a result, a test polyglot can only trigger on vulnerable pages, ensuring that non-vulnerable web applications, which constitute the vast majority, receive only the polyglot without any unintended behavior or side effects.
However, for vulnerable websites, a notification function is essential to initiate disclosure to the affected parties and improve their security.
Otherwise, our blind XSS tests would remain ``blind''.

\vspace{-1em}\paragraph{Information collected.} %
In alignment with user privacy our \ourscriptname{} (\Cref{sec:monitoring-notification}) only returns information that we use for accurate recognition and effective reporting of \blindxss{} executions in backends.
We verify instances of BXSS using the IP address and partial backend URL\@.
As part of the disclosure process, we can provide operators with both pieces of information, plus the user agent and platform strings.
The user agent information offers operators an advantage in assessing the potential impact of our reported vulnerability, as they help to distinguish between manual and automated operations.
Finally, we utilize the backend path in conjunction with the page's title as indicators of shared root-cause components, allowing us to additionally report our findings directly to the developers of these components.

\vspace{-1em}\paragraph{Candidate selection.} %
In our \emph{candidate selection} process, we adhere to the ``fairness'' principle outlined in the Menlo report by considering the top 100k Tranco domains.
With focus solely on \blindxss{}, we use a \emph{canary test} to filter out websites, like guest books, that mirror user inputs, avoiding stored and reflected XSS triggers. This test populates new forms with random tokens, subsequently checking the HTTP response and page's HTML for them. To reduce load on targets, we test each site's functionality only once, ensuring unique submissions by checking for existing duplicate header, URL, or form submissions, detailed in \Cref{sec:polyglot-transmission}.

\vspace{-1em}\paragraph{Transparency.}
For \textit{transparency} and to facilitate an \textit{opt-out} procedure, our monitoring host offers information. %
Our \ourscriptname{}'s URLs' landing page details the project,  the data we collect, and contact information for potential withdrawal from the study.
Through this channel we received two notifications from one Internet services company regarding suspicious traffic from our IP\@.
We addressed this by excluding the respective domains from subsequent visits.

Though informing thousands of operators in advance is not scalable, we ensured a \textit{vulnerability notification} was sent to the technical contacts of all affected websites.
This ensures that the underlying defects are fixed and thus exploitation is no longer possible. 
Consequently, our study's benefits in improving website security outweigh potential negatives, like polyglots causing a manual investigation. %

\vspace{-1em}
\paragraph{Design implications.} %

Although these ethical measures cannot completely eliminate the risk of a polyglot leading to a technical failure on one of the websites, we argue that the gained insights about \blindxss{} and the notification of all affected sites jointly outweigh this risk and make our study a valuable contribution for improving web security.

\smallskip{}When designing our methodology, we chose to directly collect paths and titles to identify shared root-cause components.
For instance, \Cref{tab:bxss-findings} indicates a common tool shared by backends D and F, discerned through paths.
Backends B, D and F even explicitly mention such a tool's name in their title.
We used that information to inform the respective vendors, as the vulnerability could have its root cause in a commonly used software component.

We recognize that our data collection methods, though designed with great care, are not without the risk of inadvertently capturing sensitive information in certain parts of our collected data.
Specifically, titles and paths of websites could, in theory, contain credentials, authentication tokens, or other confidential data. It is widely recognized in the field of web development and design that embedding personal data in these fields is untypical and against best practices~\citep[cf.][]{owasp-secure-design, cwe-598}.
Nonetheless, this remains a significant concern.

Our decision to collect this data was driven by the intent to notify operators of vulnerabilities in their systems, a necessary step we believed was essential for our study.
We recognize the potential pitfalls and admit that a more privacy-centered strategy could have been employed.

Our large-scale study was conducted without requesting consent.
However, we could sometimes obtain informed consent for certain aspects of the study.
Inspired by \citet{utz2023privacy}, we suggest an alternative approach for future research.
In detail, we could have abstained from collecting the backend's title and path in our initial vulnerability tests.
Then during the vulnerability disclosure, where we anticipated a limited but uncertain number of affected websites, we could have procured further information from them.
If successful, we could obtain permission to gather additional data, i.e., website titles, in a follow-up experiment, or directly ask about the utilized software components.
Such a two-fold approach would have enabled us to obtain informed consent from vulnerable entities while still learning about shared components.

Yet, this method is not without flaws.
Our study found a generally limited response rate, consistent with prior research~\cite{stock2018hear}.
Hence, this approach poses the risk of missing insights into shared root causes if no responses are received.
Nevertheless, future research might explore and evaluate this approach to determine its suitability in different contexts.

\subsection{Large-Scale Crawling Study}\label{sec:main-study}

To answer our last research question from Section~\ref{sec:rqs} (\ref{rq:3}), we perform a large-scale study on the top \num{100000} domains according to the Tranco list~\cite{le-pochat-tranco} as of Oct. 9, 2022, available at \url{https://tranco-list.eu/list/824JV}.
We use a crawler  based on the Chrome DevTools Protocol~\cite{chrome-dev-protocol} to instrument Chromium 105.0.5195.102 in headless mode.
For websites permitted by their \emph{robots.txt}, our crawler explores same-site and listed-domain links up to a depth of \num{5} or \num{30} subpages per root domain, with random link selection when needed. 
Each page receives a \num{60}\si{\s} load window; failure leads to flagging, and the crawler proceeds to another page. 
Dynamic pages receive \num{3} seconds post-\textit{load} event for pending network requests.

We conducted our study over the course of \num{20}~days using 60 parallel crawler instances on Ubuntu 20.04.5 LTS.
Our monitoring was online during this time to allow observing each submission for at least \num{2}~months.
Of the \num{1676812} visited pages, approximately 7.4\% failed to load.
Regarding the failures, about 56.1\% returned an HTTP error status code.
22.7\% of errors can be attributed to network or DNS resolution errors.
Further 15.7\% of aborted pages tried redirecting outside of the top \num{100}k domains, which we consider out of scope, or to an previously visited domain.%
The remaining 5.5\% are various errors \eg, timeouts when loading the page.

Our ethical considerations regarding deduplication and canary reduced the amount of HTML forms used for submitting our polyglots.
Collectively, these measures halved the amount of candidate forms, leaving \num{46.54}\% of identified candidates.
After respecting each input's given \inlinecode{maxlength} attribute, we submitted a total of \num{170}k forms, along with around \num{1.9}M header and \num{954}k URL submissions, equally distributed between query and path submissions.

\subsection{Uncovered BXSS Vulnerabilities}\label{sec:uncovered-vulns}
Our submissions triggered \num{20} different BXSS vulnerabilities on \num{18} websites.
In this section, we present our findings and discuss the uncovered \blindxss{} cases.
Moreover, we also discuss the efficacy of our polyglots and probing mechanisms.

\vspace{-1em}\paragraph{Vulnerable Backends.}
Regarding the given time frame and considering our initial filter (1) from \Cref{sec:bxss-filter}, we received feedback pings from \num{28} unique domains. %
After discarding stored XSS candidates with the second filter (2), we are left with \num{21} potential backend domains.
Of these \num{21} candidates, the automated part of our final filter (3) confirmed \num{8} of them to be internal websites and thus BXSS, as their URLs were unreachable \eg, because they were either private IPs, file URIs, local resources, or dotless hostnames~\cite{icann-dotless-hosts:2012}. %
Manual investigation of the remaining \num{13} reachable URLs confirmed that \num{10} of them show clear signs of \blindxss{}:
For once, \num{6} of them were login-protected pages requiring credentials or session cookies.
Another case was a web interface for a local Web Socket server that doubles as an informative website if no connection to the Web Socket port can be established, so our crawler flagged it as a public website.
Interestingly, the remaining \num{3} confirmed candidates are, to our understanding, erroneously publicly available backend pages.
Strong indicators that the pages' availability is unintended are on the one hand that they make protection-worthy data publicly available, including other visitors' IP addresses and headers, and on the other hand that their parent pages are in contrast access-protected.
Ultimately, we count these \num{10} discussed findings towards \blindxss{}, resulting in a total of \num{18} BXSS-vulnerable websites.
This demonstrates the polyglots' proficiency at uncovering \blindxss{} vulnerabilities in the backends of real websites, thus answering \ref{rq:3}.

The remaining three candidates were manually excluded and labeled as \emph{not in scope} for BXSS, because the tested website and its backend themselves were not vulnerable to our submissions, yet, we received feedback pings attributable to the polyglots submitted to these websites.
In all three cases, our \ourscriptname{} triggered in online web tools with no apparent relation to the website we tested, so we refer to this class of findings as \emph{3\textsuperscript{rd}-party XSS}.
In each case, the submitted headers were posted to evidently vulnerable online tools hours to days after we sent our submissions.
These tools include an online user-agent parser, an online XML editor and beautifier tool, as well as an online URL decoder / encoder.
We manually confirmed the three reflected XSS vulnerabilities and excluded them from our \blindxss{} results.

We disclosed our findings with all affected parties, including the previously mentioned 3\textsuperscript{rd}-party web tools.
Subsequently, we received responses in about \num{19}\% of the cases, surpassing the access-rate reported in a recent notification study~\cite{stock2018hear}.
Notably, the responses we received were entirely positive, with all respondents showing an effort to fix the issue we brought to their attention.
Moreover, we were approached by two parties requesting us to retest their fixed website, emphasizing the need for blind XSS testing strategies. %

\vspace{-1em}\paragraph{Backend Details.}
At this point, we discovered \num{18} vulnerable backends.
\Cref{tab:bxss-findings} aggregates the BXSS vulnerabilities we uncovered, showing that our findings represent websites from a wide variety of Tranco ranks, popular and less popular.
To preserve anonymity, we pseudonymized each domain in the table and shortened path and title for brevity.
In many cases, the combination of path and title are sufficient to derive the backend tools' purposes:
We observe a mix of administration tools for maintenance, management, and monitoring, as well as tools for infrastructure or business logic.
In three instances, B, D, and F, a platform name could be directly derived from the document title.
In the first case, we discovered an internal deployment of the frequently used log monitoring and reporting platform Splunk 8.2.3 and its official utility app \emph{Lookup Editor} with our polyglots. %
The other two turned out to be NetWitness Platform, a popular security information and event management tool.
We contacted both vendors to share details of our findings. %
Similar paths and titles indicate that the same software is used in the respective backends of two website pairs: D,~F and I,~K.
While the sites do not necessarily have to be connected, the location derived from their pings' IP addresses indicate proximity in both cases.

\begin{figure}[htb]
    \centering
    \includegraphics[width=\linewidth]{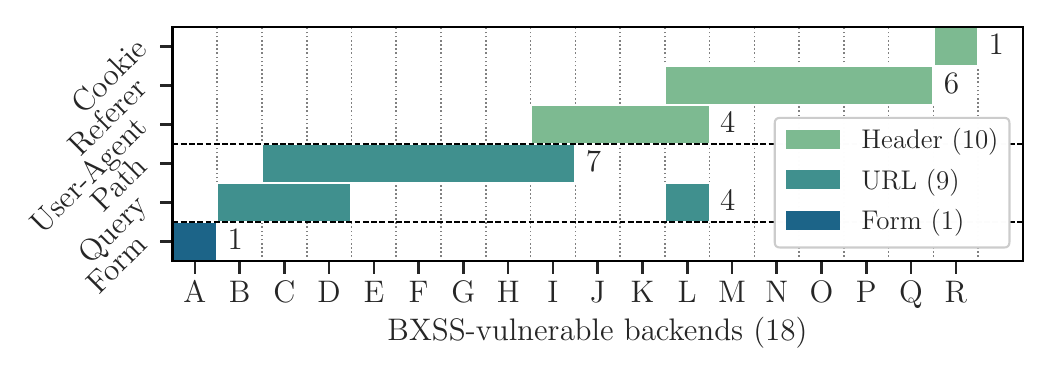}
    \caption{
    	Submission types that caused a polyglot to execute in one of the \num{18}~vulnerable backends.
    }%
    \vspace{-1em}
    \label{fig:bxss-trigger-plot}
\end{figure}

\vspace{-1em}\paragraph{BXSS Vulnerabilities.}
We count the backends based on the number of affected websites.
Since one backend can potentially have multiple BXSS vulnerabilities, we further distinguish between data flows originating from header, URL, and forms.
\Cref{fig:bxss-trigger-plot} illustrates which submission type discovered which backend.
When looking at the three groups---URL, header, and form submissions---the majority (\num{89}\%) of the backends were triggered by only one submission type.
Since the two remaining backends we discovered, I~and~L, were triggered by both URL and header submissions, we count a total of \num{20} \blindxss{} vulnerabilities on \num{18} websites.

\newcommand\mydots{\ifmmode\ldots\else\thinspace\makebox[1em][c]{.\hfil.\hfil.}\thinspace\fi}
\newcommand{\short}[0]{\textcolor{black}{[\mydots]}}
\newcommand{\timestamp}[0]{\textcolor{black}{[\emph{timestamp}]}}
\newcommand{\toolname}[0]{\textcolor{black}{[\emph{tool}]}}
\newcommand{\sitename}[0]{\textcolor{black}{[\emph{site}]}}

\newcommand{\publicdomain}[0]{$\medcircle$} %
\newcommand{\publiclockeddomain}[0]{$\circleddash$} %
\newcommand{\privatedomain}[0]{$\times$} %

\newcommand\Tstrut{\rule{0pt}{2.6ex}}   %
\newcommand\Bstrut{\rule{0pt}{2.6ex}}   %

\newcommand{\backenduri}[1]{{{#1}}}
\newcommand{\backendsymbol}[1]{{\uppercase{#1}}}

\begin{table*}[htb]
	\centering
	\tablesize
    \caption{BXSS Findings---the \num{18} backends with the corresponding website's rank, shortened path and title.}\label{tab:bxss-findings}%
	\begin{threeparttable}
        \begin{booktabs}{
            colspec={r|cclX},
            row{odd}={bg=lightgray!15},
            rows = {abovesep+=-0pt, belowsep+=-0pt},
        }
			\toprule
			\textbf{Rank} & \SetCell[c=2]{l} \textbf{Backend} & & \textbf{Backend Path} & \textbf{Backend Title} \\
			\midrule
			$90$--$100$k & \backendsymbol{a} & {\publiclockeddomain} & \backenduri{/admin/index.php} & "Welcome to service maintenance, admin!"(*)\\
			$10$--$20$k & \backendsymbol{b} & {\publiclockeddomain} & \backenduri{/ja-JP/app/lookup\_editor/lookup\_edit} & "Lookup Edit | Splunk 8.2.3"\\
			$40$--$50$k & \backendsymbol{c} & {\publicdomain} & \backenduri{/intranet/tmp/integrity-BL-KILL.html} &\\
			$90$--$100$k & \backendsymbol{d} & {\privatedomain} & \backenduri{/investigate/events} & "Investigate - NetWitness Platform"\\
			$30$--$40$k & \backendsymbol{e} & {\privatedomain} & \backenduri{/last\_one\_day\_-\_blocked\_events-\_\timestamp{}.csv.html} &\\
			$60$--$70$k & \backendsymbol{f} & {\privatedomain} & \backenduri{/investigate/events} & "Investigate - NetWitness Platform"\\
			$40$--$50$k & \backendsymbol{g} & {\publiclockeddomain} & \backenduri{/global\_administrator.aspx} & "Administrator"\\
			$20$--$30$k & \backendsymbol{h} & {\privatedomain} & \backenduri{/ajax/tst.php} &\\
			$0$--$10$k & \backendsymbol{i} & {\privatedomain} & \backenduri{/alabama\_daily\_blocks\_harding\_\_sa.\timestamp{}.csv.html} &\\
			$90$--$100$k & \backendsymbol{j} & {\publicdomain} & \backenduri{/pagestats/toonstats.php} &\\
			$90$--$100$k & \backendsymbol{k} & {\privatedomain} & \backenduri{/alabama\_daily\_blocks\_harding\_\_sa.\timestamp{}.csv.html} &\\
			$60$--$70$k & \backendsymbol{l} & {\publiclockeddomain} & \backenduri{/admin\_area/visit/v2.php}  & "Visitor Status - Daily Visitor Search | vista"(*)\\
			$80$--$90$k & \backendsymbol{m} & {\publicdomain} & \backenduri{/} & "Simple WebSocket Server -- GWSocket"\\
			$90$--$100$k & \backendsymbol{n} & {\publiclockeddomain} & \backenduri{/0xu\_x\_admin/user\_getip.asp}  & "China News Encyclopedia background management"(*)\\
			$70$--$80$k & \backendsymbol{o} & {\publiclockeddomain} & \backenduri{/\_admin/count/ip\_count.asp}  & "Access Statistics"(*)\\
			$50$--$60$k & \backendsymbol{p} & {\publicdomain} & \backenduri{/recent-referrers/} & "Referrers from past 2 days -- \sitename"\\
			$60$--$70$k & \backendsymbol{q} & {\privatedomain} & \backenduri{blank} &\\
			$70$--$80$k & \backendsymbol{r} & {\privatedomain} & \backenduri{/llurl\_fetcher\_data/f78a3c\short{}042f01.html} &\\
			\bottomrule
		\end{booktabs}
		\begin{tablenotes}
			\item \publicdomain~Backend URI is publicly available.
			\item \publiclockeddomain~Backend URI is public, but requires authentication.
			\item \privatedomain~Backend URI is unreachable. 
			\item (*) Title translated.\\
		\end{tablenotes}
	\end{threeparttable}%
    
\end{table*}

\vspace{-1em}%
\paragraph{Submission Triggers.}
Generally speaking, we observed reaction times from submission to feedback ping ranging from a few seconds for automatic processes up to \num{10} days for human interaction, with a median of around \num{6}~hours.
Regarding the geo location of machines where our polyglots triggered, we received pings from IP addresses across the world, namely Asia, Europe, America, and Africa.

Next, we further investigate each submission type in \Cref{fig:bxss-trigger-plot}.
When distinguishing between query and path submissions, we found two cases (C and D) where both types triggered in a backend.
To clarify, we do not count these as separate vulnerabilities, since these likely follow the same data flow.
Interestingly, most other URL submissions that triggered \blindxss{} were due to polyglots embedded in the path.
The three headers mainly triggered the respective backend alone, with the Referer header as the most common trigger.
Finally, in only one specific case (L), BXSS in a backend was triggered by query, User-Agent, and Referer simultaneously.

Moreover, the figure also shows that HTTP header submissions, with \num{10} cases, uncovered the most BXSS vulnerabilities, followed by URL submissions with \num{9} vulnerabilities.
Notably, forms triggered the least BXSS instances with only one occurrence.
Potential reasons for this might be the lower amount of form submissions we sent compared to headers and URLs, as well as forms being an expected input vector that may have received more attention regarding sanitization and encoding of data flowing into backends.
Despite being the commonly used attack vector~\cite{LekStoJoh13,MelDasShaBau+18,StoPfiKaiLek+15,stock2017web,BenKleBarJoh21,KleBarBen+22}, URLs still overperform as delivery type for BXSS payloads.

\begin{figure}[htb]
    \centering
    \includegraphics[width=\linewidth]{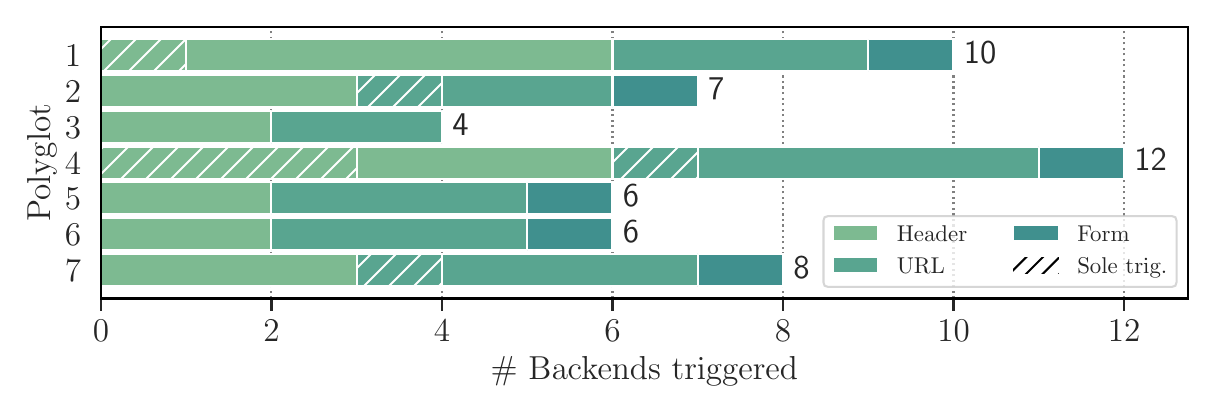}
    \caption{
    Backends triggered by the submission types. Hatched vulnerabilities were only triggered by one polyglot.
    }%
    \vspace{-1em}
    \label{fig:polyglot-trigger-plot}
\end{figure}

\paragraph{Polyglot Performance.}
As highlighted in \Cref{sec:tainting-evaluation}, manually created polyglots %
may excel in a controlled lab environment but not necessarily demonstrate comparable performance in a real-world setting.
\Cref{fig:polyglot-trigger-plot} displays the efficacy of our seven polyglots in discovering \blindxss{} vulnerabilities.
It shows that all polyglots contributed to our study's findings, with more than half of them exclusively triggering certain individual vulnerabilities.
Thus, justifying our divide-and-conquer approach based on complementing polyglots.
Overall, polyglot \num{4} was most successful regarding the number of backends triggered, as well as in the number of backends that only polyglot \num{4} was able to trigger, followed by polyglot \num{1}, and \num{7} as the top-performing polyglots.
Looking back at \Cref{fig:set-gfr-perfomance} it is interesting to see that the polyglot that was most successful in the wild is not our best on the GFR\@. %
Overall, this both highlights the need for a \emph{set of polyglots}, as well as the need for a \emph{real-world evaluation} in the wild.

\newpage 
\section{Related Work}\label{sec:related-work}%
In the following, we discuss how publications in the area of XSS detection and polyglots relate to our work.

\smallskip\textit{Stored XSS.}
While reflected XSS has been extensively studied by prior work~\cite[\eg][]{LekStoJoh13, StoPfiKaiLek+15, stock2017web, MelDasShaBau+18, BenKleBarJoh21, KleBarBen+22, son2013postman, steffens2020pmforce}, few tried to tackle the detection of \emph{stored XSS} dynamically, \ie, without access to the server-side source code.
In \citeyear{duchene2014kameleonfuzz}, \citet{duchene2014kameleonfuzz} presented KameleonFuzz, a technique to fuzz web applications guided by a genetic algorithm and a taint tracking engine.
In \citeyear{parvez2015analysis}, \citet{parvez2015analysis} analyzed the effectiveness of black-box web application scanners to detect stored vulnerabilities and found that while outperforming previous scanners~\cite{doupe2010johnny, bau2010state}, the overall detection capabilities were still quite lacking at that time.
Later in \citeyear{SteRosJohSto+19}, \citet{SteRosJohSto+19} expanded taint tracking techniques to also find Stored Client-Side XSS vulnerabilities, \ie, flows from Web Storage and cookies to dangerous sinks.
Moreover, \citet{eriksson2021black} presented BlackWidow in \citeyear{eriksson2021black}, which can discover intra-page dependencies during black-box crawling and thus uncover Stored Server-Side XSS vulnerabilities.

\textit{ML and XSS\@.}
Recently, the use of machine learning (ML), particularly reinforcement learning (RL), has gained traction in aiding XSS vulnerability detection.
In \citeyear{caturano2021discovering}, \citet{caturano2021discovering} demonstrated RL's utility in assisting human penetration testers in uncovering reflected XSS vulnerabilities. 
In \citeyear{lee2022link}, \citet{lee2022link} introduced a fully automatic RL approach, albeit limited to reflected XSS\@. 
Additionally, \citet{foley2022haxss} applied hierarchical RL to generate XSS payloads that evade the current context and bypass sanitization.

In summary, the dynamic XSS detection approaches discussed earlier share common characteristics.
They either rely on full information, employing taint tracking, are limited to Client-side XSS, or require traversing the \emph{whole} web application to identify intra-page dependencies relying on a feedback loop to guide exploitation.
In contrast, our polyglot-based approach can detect vulnerabilities \emph{blindly}, without prior knowledge or direct interactions with the vulnerable page.

\textit{XSS Polyglots.}
Some earlier work has also explored the application of polyglots and related techniques in the web context.
In \citeyear{barth2009secure}, \citet{barth2009secure} presented a PDF chameleon, which is a PostScript document that also contains some HTML, that led to XSS due to the browser's content sniffing algorithm.
In \citeyear{magazinius2013polyglots}, \citet{magazinius2013polyglots} generalized previous attacks such as the chameleon and the GIFAR~\cite{brandis2009exploring} attack under the term polyglot, and presented further attacks using PDF polyglots along with a small-scale study on 100 websites.
Additionally, various blog posts regarding universal XSS polyglots exist~\cite{ultimate-polyglot, szurek-polyglot, ostorlab-polyglot}.
High performance of the manually created Ultimate polyglot~\cite{ultimate-polyglot} on the GFR led us to include it in our baseline (see~\Cref{fig:set-gfr-perfomance}).
Mutation-based genetic algorithms~\cite{ostorlab-polyglot} constitute an interesting generation approach.
However, initial experiments showed subpar results of the published polyglots in respect to the GFR compared to the Ultimate polyglot.
This led us to not further pursue this method.
Ultimately, these blog posts lack systematic evaluation of their polyglots and, to the best of our knowledge, none of the previous publications studied the application of polyglots in the context of \blindxss{} vulnerability detection. %

\section{Conclusion}%
Our analysis sheds light on a web security problem that has stayed in the dark, so far.
While detecting and preventing XSS vulnerabilities in front end code has been a prime topic of research, little attention has been paid to analyzing these issues in backends, largely due to a lack of appropriate tools for blind security testing.
Our approach to synthesizing polyglots fills this gap and provides the basis for the first large-scale study of XSS vulnerabilities in backend code in the Web.

The generation of polyglots, however, is not only a tool for research.
Our synthesis approach is flexible.
It can be narrowed down to specific vulnerabilities as well as expanded by supplementing additional test cases.
As a result, it provides a new and versatile instrument for web security that cannot only help investigate public-to-internal flows in backends but ultimately serve as a ``Swiss-army-knife'' for general vulnerability assessment of web applications.

\section*{Acknowledgments}
We thank our shepherd and the anonymous reviewers for their valuable suggestions and comments on this paper. 
Furthermore, we would like to thank Tobias Jost, Vladislav Mladenov, and the remaining NDS team at Ruhr-Universität Bochum for their technical support, as well as Sebastian Lekies for his support with the GFR, and Angela Sasse for her guidance in our study design.
We gratefully acknowledge funding by the Deutsche Forschungsgemeinschaft (DFG, German Research Foundation) under Germany's Excellence Strategy -- EXC 2092 CASA -- 390781972, the German Federal Ministry of Education and Research (BMBF) under the project IVAN (16KIS1168), the European Research Council (ERC) under the consolidator grant MALFOY (101043410) as well as from the European Union's Horizon 2020 research and innovation programme under project TESTABLE, grant agreement No 101019206.

\section*{Availability}\label{sec:availability}

We have made our code publicly available in our companion GitHub repository at \url{https://github.com/polyxss/bxss}, encouraging further research on polyglot synthesis.

{%
\footnotesize
\def\UrlBreaks{\do\/\do-\do.\do\#}
\bibliography{paper}
}

\appendix%
\section*{Appendices}

\addcontentsline{toc}{section}{Appendices}
\renewcommand{\thesubsection}{\Alph{subsection}}

\subsection{\Ourscriptname{}}\label{appendix:probing}
\autoref{lst:probe script} shows the self-executing \ourscriptname{}, which collects minimal information about its execution environment and transmits the information via \code{XMLHttpRequest} to an API endpoint of our monitoring server.
\vspace{1em}

\begin{figure}[htb]
\scriptsize
\begin{minted}{javascript}
/* Written for ECMAScript 5.1. Please visit <redacted> for further information. Contact: <redacted> */

(function () { 
	/* collect minimal information */
	var data = {
		"id": "info",
		"title": document.title,
		"protocol": document.location.protocol.replace(":", ""),
		"domain": document.domain,
		"port": document.location.port,
		"pathname": document.location.pathname,
		"navigator_ua": window.navigator.userAgent,
		"navigator_platform": window.navigator.platform
	};

	/* report */
	var url = "https://<redacted>/callback?";	
	for (var key in data) {  
		url += key + "=" + encodeURIComponent(data[key]) + "&";
	}
	url += "timestamp=" + (new Date().getTime()).toString();

	var xhr = new XMLHttpRequest();
	xhr.open('GET', url, true);
	xhr.send();
})();
\end{minted}

\caption{\Ourscriptname{}}%
\label{lst:probe script}
\end{figure}

\vspace{-1em}
\subsection{Data Management}
\label{appendix:data-management}
Due to our notification script's design, as detailed in \Cref{sec:monitoring-notification} and \Cref{appendix:probing}, we received only a minimal amount of data.
Notably, the backend URL's query and fragment were never transmitted.
We expected to receive no personally identifiable information (PII) in either the path or the title of affected webpages.
However, we prepared a data management strategy to handle potential PII transmissions from our \ourscriptname{}'s feedback pings.
Our strategy entails to manually replace potentially received PII with placeholders.
Fortunately, as shown in \Cref{tab:bxss-findings} neither the backend paths nor titles contained such information.
IP addresses have a distinctive role, as they are indirectly received via a feedback ping from a vulnerable backend.
Initially used to validate instances of \blindxss{} (\Cref{sec:bxss-filter}), they later assist in meaningful disclosure and forensic analysis.
Supplying operators with both the IP address and user agent information, facilitating distinction between manual actions and automated processes, helps them assess the vulnerability's impact.
Post-disclosure, all retained IP addresses and user agents were deleted.
Ultimately, the data this paper is based on will be archived as scientific evidence according to our institution's guidelines.%

\subsection{Alternative Generation Approaches}%
\label{appendix:alternative-generation}

Our polyglot synthesis for BXSS with MCTS has been successful in generating a complementary polyglot set. As our generation approach is agnostic to this algorithm, however, we investigate three alternative algorithms that could also be applied to our setting for constructing polyglots: \emph{random selection}, \emph{greedy selection}, and \emph{Q-learning}. %

\smallskip
\emph{Random selection ---} This method leverages our MCTS implementation, as depicted in Algorithm \ref{alg:gen-polyglots}, but incorporates random selection and a random playout phase. In the selection phase, each child is given an equal probability, and one is selected at random. Once the end condition---specifically, the maximum payload length---is met, the constructed polyglot undergoes evaluation on the small testbed.

\smallskip
\emph{Greedy selection ---} This method generates a polyglot by continuously appending the best next token.
Since we cannot know in advance which token is the best, we probe each token by appending the respective token to the current polyglot and evaluating the resulting polyglot on the testbed, thus implementing a greedy search.
If multiple tokens achieve the same performance, we select one randomly to append.
In contrast to \emph{MCTS} and random selection, the greedy method evaluates unfinished polyglots in order to choose the next token.

\smallskip
\emph{Q-learning ---} This method builds on \emph{Q-learning}~\cite{watkins1992q}, a popular reinforcement learning technique.
The method populates a table of state-action pairs $Q(s_t, a_t)$ to determine the best action $a_t$ in a given state $s_t$. 
Algorithm \ref{alg:q-learning} shows our \emph{Q-learning} implementation. 
It first chooses either a random action or the best next action and then evaluates the resulting state on the testbed, updates the q-values and saves the polyglot if it performed best.
For \emph{Q-learning} we set the learning rate $\alpha=0.1$ and the discount factor $\gamma=0.99$ as well as the simulated annealing parameters: initial exploration probability $p=1$, minimal exploration probability $p_{min}=0.01$ and exploration decay $\Gamma=0.95$.

\begin{algorithm}[htb]
\scriptsize
\DontPrintSemicolon{}
\SetKwInput{KwInput}{Input}
\SetKwInput{KwOutput}{Output}
\SetAlgoNlRelativeSize{-2} %
\SetNlSty{normalfont}{}{} %
\KwOutput{polyglot string}
$s_{best} \gets$ null \;
$r_{best} \gets 0$ \;
$s_0$ $\gets []$  \;
\While{evaluation budget is not exceeded}{
    \For{$t \gets 0$; polyglot not complete; $t \gets t+1$}{
        \If{$rand(0, 1) < p_{exp}$} {
            $a_t$ $\gets$ choose\_random($s_t$);
        } \Else {
            $a_t \gets \max_a Q(s_t, a)$ \;
        }

        $s_{t+1} \gets s_t + [a_t]$ \;
        $r_t \gets$ evaluate($s_{t+1}$) \;
        $Q(s_t,a_t) \gets Q(s_t,a_t) + \alpha(r_t+\gamma \max_a Q(s_{t+1}, a) - Q(s_t, a_t))$ \;

        \If{$r_{t} > r_{best}$} {
            $r_{best} \gets r_t$ \;
            $s_{best} \gets s_{t+1}$ \;
        }

        $p \gets max(\Gamma \cdot p, p_{min})$  \;
    }
}
\KwRet~$s_{best}$ \;
\caption{Generating a polyglot with Q-learning}\label{alg:q-learning}
\end{algorithm}

\vspace{-1em}\paragraph{Comparative evaluation.}
We compare the approaches based on the coverage they achieve on the GFR and the number of polyglots they require to reach that coverage. The experiment consists of generating a set of \num{10} polyglots with each algorithm, evaluating the resulting sets on the GFR and removing polyglots from the sets that do not contribute to the overall performance. To account for randomness, the experiment is repeated \num{10} times for each approach.

In the generation phase, each approach iteratively generates polyglots on the small testbed. 
After each iteration, we remove tests that are covered by the polyglots. 
Since calls to the testbed are computationally the most expensive component of each algorithm, all methods are given a budget of \num{12,000} evaluation calls to the testbed during each iteration.

\begin{figure}[htb]
    \centering
    \includegraphics[width=\linewidth]{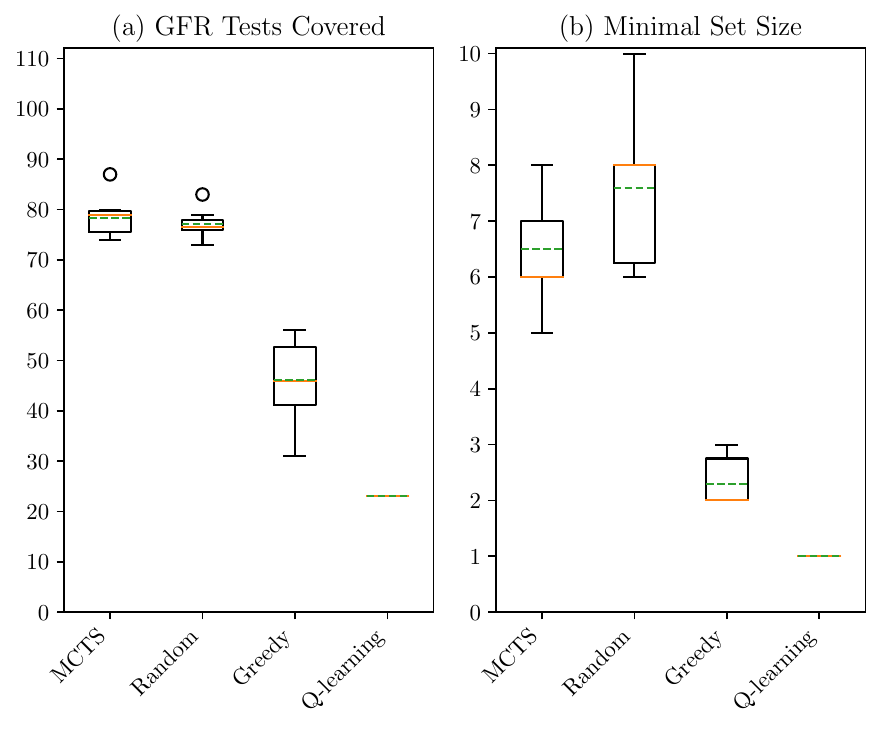}
    \caption{
    MCTS, Random, Greedy and Q-learning comparison; each generating $10$ polyglots in $10$ parallel runs
    }\label{fig:comparison-approaches}
\end{figure}

The evaluation of our comparative experiment's results are shown in \Cref{fig:comparison-approaches}.
Plot (a) aggregates the resulting set coverage on the GFR of the \num{10} parallel repetitions of each approach as boxplots. 
Plot (b) displays boxplots of the set sizes the same approaches achieved.
Generally, smaller polyglot set sizes are preferred because they would result in fewer requests being sent to a system under test when testing for client-side XSS (\Cref{sec:taint-validation}) or \blindxss{} (\Cref{sec:bxss-study}). %
While Q-learning and greedy selection produce smaller polyglot sets, their overall coverages on the GFR are significantly lower than those of MCTS and random selection.
We can therefore discard both of them as alternative polyglot synthesis approaches.
In terms of coverage, the random method performs only slightly worse than MCTS on average.
However, MCTS consistently achieves a lower set size than random selection.
We believe this is the result of \emph{MCTS}'s knowledge aggregated over multiple games, which allows it to generate more powerful polyglots. %

\subsection{Google Firing Range}\label{appendix:gfr-exclusions}
\newcommand{\gfrcategory}[1]{``{#1}''}

In \Cref{sec:polyglot-evaluation} we discuss the construction of our comprehensive XSS testbed, which was derived from a subset of the GFR~\cite{github-firing-range} tests to determine the efficacy of the polyglots.
This section details which tests were excluded and why.

\definecolor{OliveGreen}{HTML}{3C8031}
\definecolor{BrickRed}{HTML}{B6321C}
\newcommand{\inscope}[0]{\textcolor{OliveGreen}{\ding{51}}}
\newcommand{\noscope}[0]{\textcolor{BrickRed}{\ding{55}}}

\begin{table}[htb]
	\centering
	\tablesize
	\begin{threeparttable}
		\caption{Vulnerability categories of the GFR 0.48 and their general applicability for our \blindxss{} testbed}\label{tab:appendix:gfr-categories}
		\begin{tabularx}{\linewidth}{rXcl}
			\toprule
			\textbf{\#} & \textbf{Category} & \textbf{Scope} & \textbf{Reasoning} \\
            \midrule
            1 & Address DOM XSS & \inscope & XSS-related \\
            2 & Angular-based XSSes & \noscope & framework-specific \\
            3 & Bad JavaScript imports & \noscope & non-XSS-related \\
            4 & CORS related vulnerabilities & \noscope & non-XSS-related \\
            5 & DOM XSS & \inscope & XSS-related \\
            6 & Escaped XSS & \inscope & XSS-related \\
            7 & Flash Injection & \noscope &  non-XSS related \\
            8 & Mixed content & \noscope & non-XSS related \\
            9 & Redirect XSS & \inscope & XSS-related \\ 
            10 & Reflected XSS & \inscope & XSS-related \\
            11 & Remote inclusion XSS & \inscope & XSS-related \\
            12 & Reverse ClickJacking & \noscope & non-XSS related \\
            13 & Tag based XSS & \inscope & XSS-related \\
            14 & URL-based DOM XSS & \inscope & XSS-related \\
            15 & Vulnerable libraries & \noscope & non-XSS related \\
            16 & Leaked httpOnly cookie & \noscope & non-XSS related \\
            17 & Invalid framing configuration & \noscope & non-XSS related \\
            \bottomrule
		\end{tabularx}
		\begin{tablenotes}
			\item \inscope{} category is generally in scope\hspace{2em}\noscope{} category is out of scope
		\end{tablenotes}
	\end{threeparttable}%
\end{table}

\newcommand{\myarrow}{\textcolor{red}{$\hookrightarrow$}}
\soulregister{\myarrow}{1}
\sethlcolor{white}
\newcommand\gfrpath[1]{\hl{\texttt{#1}}}  %

\begin{table}[htb]
	\centering
	\tablesize
    \caption{Excluded GFR tests from the categories in scope, referenced by path}\label{tab:appendix:gfr-excluded-tests}%
    \begin{booktabs}{
        colspec={X},
        row{odd}={bg=lightgray!25},
        rows = {abovesep+=2pt, belowsep+=2pt},
    }
    \toprule
    
    \textbf{No solution confirmed} (20) \\ 
    \gfrpath{/dom/toxicdom/postMessage/improperOriginValidationWithPartialStringComparison}, \gfrpath{/dom/toxicdom/postMessage/improperOriginValidationWithRegExp}, \gfrpath{/dom/javascripturi.html}, \gfrpath{/escape/serverside/encodeUrl/tagname}, \gfrpath{/escape/serverside/encodeUrl/js\_assignment}, \gfrpath{/escape/serverside/encodeUrl/js\_eval}, \gfrpath{/escape/serverside/escapeHtml/attribute\_script}, \gfrpath{/escape/serverside/escapeHtml/href}, \gfrpath{/escape/serverside/encodeUrl/href}, \gfrpath{/tags/tag}, \gfrpath{/tags/tag/div}, \gfrpath{/tags/tag/img}, \gfrpath{/tags/tag/style}, \gfrpath{/tags/tag/iframe}, \gfrpath{/tags/tag/div/style}, \gfrpath{/tags/tag/a/href}, \gfrpath{/tags/tag/a/style}, \gfrpath{/tags/tag/script/src}, \gfrpath{/tags/tag/body/onload}, \gfrpath{/urldom/location/hash/script.src.partial\_query} \\
    
    \textbf{Technical reasons} (5) \\ 
    \gfrpath{/dom/toxicdom/document/referrer/eval}, \gfrpath{/dom/toxicdom/document/referrer/innerHtml}, \gfrpath{/dom/toxicdom/document/referrer/documentWrite}, \gfrpath{/stored/index.html}, \gfrpath{/urldom/location/hash/window.open} \\
    
    \textbf{Solution too narrow} (12) \\ 
    \gfrpath{/dom/toxicdom/postMessage/documentWrite}, \gfrpath{/reflected/url/href}, \gfrpath{/reflected/url/script\_src}, \gfrpath{/remoteinclude/parameter/script}, \gfrpath{/remoteinclude/script\_hash.html}, \gfrpath{/urldom/location/hash/base.href}, \gfrpath{/urldom/location/hash/fetch}, \gfrpath{/urldom/location/hash/script.href}, \gfrpath{/urldom/location/hash/script.src}, \gfrpath{/urldom/location/hash/xhr.open}, \gfrpath{/urldom/location/hash/script.src.partial\_domain}, \gfrpath{/urldom/location/hash/script.src.partial\_path} \\
    
    \textbf{Technology out-of-scope} (8) \\ 
    \gfrpath{/remoteinclude/parameter/object\_raw}, \gfrpath{/remoteinclude/object\_hash.html}, \gfrpath{/urldom/location/hash/embed.src}, \gfrpath{/urldom/location/hash/object.data}, \gfrpath{/urldom/location/hash/param.code.value}, \gfrpath{/urldom/location/hash/param.movie.value}, \gfrpath{/urldom/location/hash/param.src.value}, \gfrpath{/urldom/location/hash/param.url.value} \\
    
    \bottomrule
    \end{booktabs}
\end{table}

The GFR is structured as a crawlable list of subpages covering different categories of web vulnerabilities.
Each category produces tests from a mix of sink, source, and countermeasures.
Due to this setup, certain combinations result in unsolvable tests.
To create our comprehensive and solvable XSS testbed, it is essential to first filter the GFR test cases. 

\smallskip
To begin with, we omitted all test categories unrelated to XSS\@.
As outlined in \Cref{tab:appendix:gfr-categories}, the categories \gfrcategory{Bad JavaScript imports}, \gfrcategory{CORS related vulnerabilities}, \gfrcategory{Flash Injection}, \gfrcategory{Mixed content}, \gfrcategory{Reverse Clickjacking}, \gfrcategory{Vulnerable libraries}, \gfrcategory{Leaked httpOnly cookie}, and \gfrcategory{Invalid framing configuration} are out of scope as they are not related to XSS vulnerabilities. 
Additionally, we disregarded the \gfrcategory{Angular-based XSSes} category given its focus on a particular framework, AngularJS, which is beyond our study's scope.

From the categories that remained relevant post our preliminary filtering, \Cref{tab:appendix:gfr-excluded-tests} provides a list of the excluded tests alongside reasons for their omission.
For brevity, tests are identified by their path, accessible by appending the test path to GFR's main URL {\small{}\url{public-firing-range.appspot.com}}.

Our subsequent filtering entailed the removal of tests lacking solutions.
Some tests became unsolvable due to the combination of sinks and countermeasures, modern browser security features, or obsoleted features.
Fortunately, we acquired a list of solvable GFR tests from Google.
After manual confirmation, we removed tests without a solution.
Secondly, a handful of tests were removed for technical reasons, such as being removed from the GFR (Stored XSS), 
or tests involving multiple windows.
Tests with exceptionally restrictive solutions, out-of-scope for a polyglot, were also ruled out, including those with stringent filters, or those solvable using only a URL\@.
Likewise, we omitted tests exceeding our technology boundaries, such as those using Adobe Flash, Base64, and SVG\@.
This leaves us with \num{111} tests.

The GFR tests involve various input sources, including form submissions, URL parameters, or \emph{PostMessages}.
Some demand clicks or page reloads post-input.
Using Puppeteer, our test software provides a polyglot to each relevant GFR test through the suitable input method, meeting the post-submission requisites.
It then waits for an XSS success signal through a specific log message.
We use Puppeteer cluster to test one polyglot on multiple tests in parallel.
After a run concludes, the polyglot's result in each test are returned.
The implementation of our testing approach is published in our companion repository.

\end{document}